%% file: main.tex
\documentclass[sigconf]{acmart}

\input{packages}

\AtBeginDocument{%
  }

\copyrightyear{2026}
\acmYear{2026}
\setcopyright{cc}
\setcctype{by}
\acmConference[CHI '26]{Proceedings of the 2026 CHI Conference on Human Factors in Computing Systems}{April 13--17, 2026}{Barcelona, Spain}
\acmBooktitle{Proceedings of the 2026 CHI Conference on Human Factors in Computing Systems (CHI '26), April 13--17, 2026, Barcelona, Spain}
\acmPrice{}
\acmDOI{10.1145/3772318.3790460}
\acmISBN{979-8-4007-2278-3/2026/04}

\begin{document}

\title[Norms and Observer Effects in Shared LLM Accounts]{“Don’t Look, But I Know You Do”:\texorpdfstring{\\}{ }%
Norms and Observer Effects in Shared LLM Accounts}

\author{Ji Eun Song}
\orcid{0000-0002-3118-0599}
\email{jieun.song@snu.ac.kr}
\affiliation{%
  \department{GSCST}
  \institution{Seoul National University}
  \city{Seoul}  
  \country{Republic of Korea}
}

\author{Eunchae Lee}
\affiliation{%
  \institution{KAIST}
  \city{Daejon}
  \country{Republic of Korea}
  }
\email{chaelee@kaist.ac.kr}

\author{Juhee Im}
\affiliation{%
  \department{GSCST}
  \institution{Seoul National University}
  \city{Seoul}
  \country{Republic of Korea}
  }
\email{jhdsny1105@snu.ac.kr}

\author{Hyunsoo Jang}
\affiliation{%
  \department{GSCST}
  \institution{Seoul National University}
  \city{Seoul}
  \country{Republic of Korea}
  }
\email{gnfk0119@snu.ac.kr}

\author{Eunji Kim}
\affiliation{%
  \department{GSCST}
  \institution{Seoul National University}
  \city{Seoul}
  \country{Republic of Korea}
  }
\email{is_eunji@snu.ac.kr}

\author{Joongseek Lee}
\affiliation{%
  \department{GSCST}
  \institution{Seoul National University}
  \city{Seoul}
  \country{Republic of Korea}
  }
\email{joonlee8@snu.ac.kr}

\renewcommand{\shortauthors}{Song et al.}

\begin{abstract}
Account sharing is common in subscription services and is now extending to generative AI platforms, which are still primarily designed for individual use. Sharing often requires workarounds that create new tensions. This study examines how LLM subscriptions are shared and the norms that develop. We combined a survey of 245 users with interviews of 36 participants to understand both patterns and lived experiences. Our analysis identified four types of account sharing, organized along two dimensions: whether the owner uses the account and whether subscription costs are shared. Within these types, we examined how norms were formed and how their fragility, especially privacy, became evident in practice. Users, fully aware of this, subtly adjusted their behavior, which we interpret through the lens of the observer effect. We frame LLM account sharing as a social practice of appropriation and outline design implications to adapt single-user platforms to multi-user realities.
\end{abstract}



\maketitle

\input{sections/01-introduction}
\input{sections/02-relatedwork}
\input{sections/03-methodology}

\input{sections/04-findings}
\input{sections/05-discussion}
\input{sections/06-limitations}
\input{sections/07-conclusion}

\bibliographystyle{ACM-Reference-Format}
\bibliography{references/bibliography}

\appendix
\input{sections/Appendix-A}

\input{sections/Appendix-B}

\end{document}

%% file: packages.tex
\usepackage{multirow}

\usepackage{float}

\usepackage{soul}      
\usepackage{xcolor}    
\sethlcolor{yellow}    

\usepackage{natbib}   

\soulregister\cite7   

%% file: sections/01-introduction.tex
\section{INTRODUCTION}

Account sharing has become increasingly widespread across subscription services, including streaming and cloud platforms. Families share streaming accounts to save money \cite{save-streaming}, teams use cloud storage collectively, and even strangers split costs through dedicated platforms \cite{gamsgo}. Large language model (LLM) services are no exception. Major providers like OpenAI reveal this tension: they explicitly prohibit account sharing in their terms of service \cite{openaisharingpolicy} yet offer team plans that only partially address multi-user needs \cite{openaibusiness}.

LLM services center on individual user profiles. This ensures consistent personalization for each user but creates problems when multiple people share the same account. As LLMs become increasingly common in education, collaboration, and creative work, a single account often serves as a shared resource. Through sharing, users appropriate these systems, transforming single-user designs into multi-user environments \cite{Schmidt1992}. Understanding this process is crucial to seeing how LLMs shift from personalized tools to collaborative and relational resources.

In this study, we approach LLM account sharing as a form of appropriation—users recontextualizing individually designed systems for multi-user situations. As users collaborate to reconfigure these systems \cite{Boardman2004PIM}, they develop practices that rely on explicit rules and more on implicit norms. Therefore, appropriation emphasizes not only how people adapt technologies but also how they transform shared resources into functional social orders. This expands long-standing CSCW discussions of coordination into a new technological realm while guiding the design of multi-user environments.

Unlike streaming services or collaborative coding tools, shared LLM accounts leave behind linguistic traces that reveal not only what was accessed but also how people think, reason, and feel. These "cognitive traces" include process-level artifacts that show how a problem is framed and solved, including context, intermediate reasoning, affective cues, and strategies, making the problem-solving process visible, reusable, and socially significant. While Netflix histories or GitHub commits record specific outputs, LLM conversations preserve dialogic processes—questions, tentative reasoning, problem-solving strategies, and emotional expressions. This makes LLM account sharing fundamentally different: users reveal not just their usage patterns but also fragments of their cognition. 

This study investigates how LLM subscriptions become shared resources and what social dynamics emerge. Specifically, we address three research questions:
\begin{itemize}
    \item \textit{RQ1:} In what forms and under what conditions does LLM account sharing occur?
    \item \textit{RQ2:} How are implicit norms formed and interpreted in these shared practices?
    \item \textit{RQ3:} How do the reasoning traces of LLM environment shape implicit norms enacted in shared accounts due to its nature?
\end{itemize}
We employed a mixed-methods approach: a survey with 245 participants followed by semi-structured interviews with 36 participants. 

Our analysis identified four distinct types of account sharing, structured by whether the account owner actively used the service and whether subscription costs were shared among the account members. Across all four types, in the absence of explicit rules, users interpreted the norms differently, resulting in divergent experiences. On the one hand, users revealed themselves through visible traces; on the other, they constrained their behavior in consideration of others' awareness.

This paper makes three main contributions. First, we identify a 2×2 typology of LLM account sharing (owner participation × cost-sharing) and show how it structures motivations and coordination logics, characterizing LLM account sharing as technological appropriation that transforms individuals tools into social resources. Unlike prior work on account sharing in streaming and enterprise systems, this typology explicitly accounts for reasoning-trace visibility as a key coordination concern. Second, we reveal how implicit norms emerge through shared traces and how their inconsistent interpretations create observer-effect tensions, extending classic concepts of awareness and social translucence by focusing on linguistic reasoning traces rather than behavioral logs. Third, based on our research we derive design strategies for negotiation tools, social translucence, and graded visibility that adapt single-user LLM interfaces to multi-user realities. We derive design strategies for negotiation, translucence, and visibility gradation that concretely address this tension in LLM interfaces.

%% file: sections/02-relatedwork.tex
\section{RELATED WORK}

To understand LLM account sharing as a socio-technical phenomenon, we draw on three theoretical streams: appropriation theory, implicit norm formation and observer effects in HCI. Appropriation theory helps us treat account sharing as design-in-use rather than mere rule-breaking, clarifying how single-user systems become shared resources (RQ1). Work on implicit norms explains how unspoken “house rules” emerge, travel, and break down when people adapt technologies together (RQ2). Research on observer effects in HCI shows how visibility and the prospect of being watched reshape everyday digital behavior, which is central when shared LLM histories expose rich traces of users’ reasoning and self-presentation (RQ3). Together, these perspectives ground our analysis of sharing types, norm dynamics, and observer-effect tensions in shared LLM accounts.

\subsection{Account Sharing as Appropriation}

Appropriation refers to how users adopt and adapt technologies, beyond their intended design, integrating them into their practices \cite{Stevens2009AppropriationInfrastructure}. Classic CSCW and HCI studies demonstrate how users routinely appropriate systems in unanticipated ways, suggesting that design should accommodate such reconfigurations \cite{Dourish2003}. These perspectives are located within broader HCI theory on contemporary “design-in-use” \cite{Salovaara2006StudyingAppropriation}, everyday design in the home \cite{Wakkary2008AspectsOfEverydayDesign}, and groupware \cite{BanslerHavn2006Sensemaking}. 

From this perspective, account sharing is not just a breaking of rules, but a form of appropriation. Traditional HCI research has primarily focused on single-user interfaces, assuming individual ownership and use of accounts \cite{Rogers2011HCITheory}. However, in real life, people naturally share technologies, creating a disconnect between individual-focused design and multi-user practices \cite{ObadaObieh2020Burden}.

To address such forms of appropriation, earlier HCI research proposed various multi-access solutions. Recent efforts aim to bridge this gap by developing features like multi-profile interfaces, partial access via app-specific unlocks, or state-based access models that extend beyond simple lock and unlock mechanisms \cite{Hayashi2012Goldilocks}. Yet these solutions have not prevented continued account appropriation \cite{Song2021IMightBeUsingHis}. 

Prior research identifies four primary motivations for account sharing. Convenience drives sharing among family members who value seamless access \cite{FrohlichKraut2003SocialContext}. Trust motivates couples to share accounts as symbolic intimacy \cite{Blythe2013Circumvention, Jacobs2016CaringSharing, Matthews2016EverydayDeviceSharing}. Collaboration needs lead items to share accounts despite security policies, centralizing work and reducing coordination overhead \cite{AdamsSasse1999Users, Koppel2015Workarounds, Song2019NormalEasy}. Finally, economic factors increasingly drive sharing as subscription costs rise \cite{Prottoy2025IfWeHadTheOption, Song2021IMightBeUsingHis}.

These appropriation practices, whether motivated by convenience, trust, collaboration, or economics, rarely remain unstructured. In the absence of explicit system support, users develop informal “house rules” to coordinate shared use, deciding who may log in, what content can be accessed, and how conflicts are resolved. These emergent arrangements show how appropriation and norm-making are deeply intertwined, yet because the rules are unwritten and socially enforced, they remain fragile and prone to breakdown. Building on these ideas, we treat shared LLM accounts as a concrete instance of appropriation in which users retrofit single-profile services into multi-user infrastructures. In this way, four types of LLM sharing recurrently emerge, each with distinct motivations and coordination logics. By mapping these types and their motivations, we extend traditional practices of account sharing in streaming services and workplace systems to LLM subscriptions. Here, appropriation also emerges, crucially, however, it is much more complex. Not only do these forms of sharing arise in relation to logins and passwords, but also the dynamics of chat history and cognitive traces. These added burdens create new demands for coordination, leading us to examine how implicit norms emerge in these settings.

\subsection{Implicit Norm and its Vulnerability}

Implicit norms refer to the unspoken rules that govern how people coordinate their expectations and conduct themselves without formal policies or explicit agreements. 

Arising through use, implicit norms are strongly linked to appropriation. Research on appropriation shows that users reinterpret technologies beyond designers' original intentions, and this reorganization occurs alongside informal “house rules” that support shared practices \cite{Salovaara2006StudyingAppropriation, Salovaara2011EverydayAppropriations}. Community studies highlight the same concept on a larger scale: newcomers learn implicit norms by observing and emulating more experienced members, and repeated practice helps reinforce these expectations \cite{Bryant2005BecomingWikipedian}. Over time, some unwritten rules become explicit guidelines \cite{Butler2008WikipediaPolicies}. However, many remain tacit and are maintained through social enforcement \cite{Fiesler2019FanNorms}. In live, leaderful environments, example-setting by high-status members accelerates adoption: others imitate visible role models, turning patterns into norms without formal instruction \cite{Seering2017TwitchNorms}.

This unwritten nature makes implicit norms inherently fragile. First, they are community-specific and often opaque to outsiders; participating across communities increases the risk of accidental violations \cite{Chandrasekharan2018RedditNormViolations}. Second, enforcement is socially costly: people prefer subtle, face-saving sanctions over confrontation, which leaves some violations unaddressed \cite{Rashidi2020SanctioningNorms}. Third, norms are context-sensitive: negative mood or exposure to norm-breaking content can trigger cascades of further violations, showing how adherence depends on situational cues \cite{Cheng2017Troll}. Fourth, technical shifts can destabilize previously effective norms or create unconventional ones \cite{Bernstein2011Fourchan}. Finally, cultural drift and value conflict—especially with newcomer influxes—can erode safety-critical norms, creating  risks when these invisible protections fail\cite{Dym2020NormVulnerability}. 

Privacy practices illustrate a situation where implicit norms both create and undermine boundaries. In close relationships, individuals jointly manage disclosure through implicit norms, working together to set boundaries around shared information \cite{Lampinen2011DisclosureManagement}. When sharing accounts and devices, implicit privacy norms dictate who can see or use what, when, and how; these norms often develop as couples, families, or teams adapt single-user systems for multiple users \cite{Park2018ShareAndShareAlike}. The risks are evident: users often adopt unsafe practices (e.g., sending passwords via text, creating sharing lists), and delay explicit conversations to avoid awkwardness, and find it difficult to understand when relationships or roles change \cite{ObadaObieh2020Burden}.

Appropriation and rule-making are inseparable: as users bend systems to fit local needs, they simultaneously mint and teach norms that make the bending sustainable.

Drawing on prior literature, we trace how implicit norms around privacy, boundaries, and access are formed—and when and how they begin to fracture, within the four sharing types we identify. Many “house rules” remain individually imagined rather than collectively negotiated, producing gaps in expectations even among close ties. These misalignments yield a paradox of privacy enforcement: checking whether others are following the rules can itself require the need to view other threads which the very rules forbid. Taken together, these dynamics show how shared LLM accounts can appear to function smoothly in some situations yet become sites of tension and quiet norm breakdown in others. To further unpack this visibility–behavior interplay, we next turn to the observer effect, which highlights how the presence and gaze of others reshape everyday digital practices.

\subsection{Observer Effect in HCI}

The observer effect —sometimes called the Hawthorne effect \cite{ChiesaHobbs2008Hawthorne} — refers to how observation necessarily changes the observed phenomenon \cite{Baclawski2018ObserverEffect}. While the Hawthorne effect traditionally refers to behavioral changes in experimental settings \cite{CoombsSmith2003HawthorneEffect, Fox2008ClinicalEstimationFetalWeightHawthorne, LeCompte1982ReliabilityValidity, Oswald2014HandlingHawthorne}, we adopt the broader observer effect concept for everyday digital environments \cite{Obrenovic2014HawthorneHCI}. However, in this study, we define it as behavioral changes resulting from the presence and gaze of others during everyday digital use. 

This approach begins with a key question about how it differs from concepts of awareness and visibility in HCI. In CSCW, awareness highlights how systems show who, what, and when to assist with coordination, while visibility is argued to encourage collaboration and accountability. However, the observer effect centers on how this visibility extends beyond sharing information, actually altering behaviors. For example, Birnholtz et al. \cite{Birnholtz2012Visibility} found that whether a partner knew they were being “checked on” significantly affected how often participants monitored them. Visible monitoring boosted accountability and reduced unnecessary spying, while invisible monitoring led to more voyeuristic behaviors. 

In HCI, increasing visibility has been argued to promote collaboration and accountability \cite{Erickson2000SocialTranslucence}. However, such designs can also increase users’ awareness of surveillance, leading to self-censorship and impression management \cite{Mitchell2011ImpressionManagementHCI}. Perceived observation often results in self-censorship, self-presentation, or behavioral inhibition. Studies on Facebook and Twitter show that users frequently write posts but hesitate to publish them once they consider who might see them—an act of last-minute self-censorship \cite{Sleeper2013FacebookCensorship, Das2013SelfCensoring, Marwick2011ImaginedAudience}. Context collapse worsens this issue: users imagine their most judgmental audience member and modify their behavior accordingly \cite{Marder2016ChillingFacebook}. In real-world settings, interactions with public chatbots vary depending on the audience: with friends nearby, users joke or show off; with strangers, they stay formal; alone, they are more direct \cite{Candello2019AudienceEffectCHI}. These patterns reflect social facilitation theory, which posits that simply having others present can influence performance and expression.

Recent studies show the observer effect directly in HCI settings. On transparent collaboration platforms like GitHub, users are more careful with commit messages and contributions because they know their work is publicly visible \cite{dabbish2012social}. On social media, participants in a research study altered their posting behavior after being informed their data was being observed, demonstrating a Hawthorne-like effect \cite{Saha2024ObserverEffect}.

While the term “observer effect” is rarely explicit in HCI literature, studies consistently show how users develop workarounds when aware of others' presence. Beyond couples, Netflix users reported avoiding certain shows to prevent others from viewing their watch history \cite{Sailaja2022AccountSharing}. A qualitative study by Sambasivan et al. \cite{Sambasivan2018Privacy} documented how women in India, Pakistan, and Bangladesh protect their privacy in shared phone environments by locking apps, routinely deleting records, or avoiding using the device in front of others. Users also manage visibility by separating platforms. In collectivist cultural contexts with strong religious or traditional norms, women configured accounts to limit exposure to male audiences by selectively sharing content on specific platforms \cite{Naveed2022PrivacyBeyondWEIRD}. In extreme cases of intimate partner abuse, survivors created secret email or social media accounts to maintain hidden networks of support. One respondent reported: “My ex-husband logged into my Google account to track my location, so I turned GPS off completely.” \cite{Matthews2017StoriesFromSurvivors} 

 Such analysis shows that the observer effect directly correlates to how users interact with the digital world. As shown, users often use workarounds to protect their privacy and prevent external judgments. However, little is known about how these dynamics play out in shared LLM accounts where the production of chats within an LLM expose the users' reasoning as both a learning resource and a vulnerability. Taken together, these three theoretical lenses that have been discussed allow us to treat shared LLM accounts as sites where appropriation, implicit norm‑making, and observer effects intersect, motivating our empirical mapping of sharing types (RQ1) and norm dynamics (RQ2) and our theorization of observer‑effect tension around reasoning traces (RQ3).

%% file: sections/03-methodology.tex
\section{METHOD}

To understand users' strategies for sharing LLM accounts and the resulting social and psychological effects, we employed a mixed-methods approach. The study protocol was approved by our institution's human subjects review board (IRB). We first conducted a quantitative survey of 245 participants to examine the prevalence of account sharing and its associated norms. Subsequently, we conducted semi-structured interviews with 36 participants to explore more subtle dynamics including tacit knowledge, social understanding, and boundary negotiation. This sequential approach allowed us to combine broad patterns from the survey with detailed insights from the interviews. 

As Blom \cite{Blom2005ContextualCultural} suggests, focusing on a specific cultural context can help identify common behavioral patterns. Building on Hofstede’s cultural dimension of individualism versus collectivism \cite{Hofstede2011Dimensionalizing}, we situate our sample within a collectivist setting. While our findings are culturally situated, they provide insights into how account sharing manifests in collectivist contexts and how cultural values shape coordination practices and observer effects. 

\begin{table}[]
    \centering
    \caption{Overview of interview participants (N=36), including account role (Owner=O, Member=M, Lendee=L), platform(s) used, length of subscription use, and the number of co-users recognized by each participant.}
    \label{tab:table1}
    \input{tables/table1}

\end{table}

\subsection{Survey}
The survey was designed to investigate overall patterns of account sharing (RQ1) and users’ perceptions of implicit norms (RQ2). It took approximately ten minutes to complete and included questions about sharing experiences, cost arrangements, and open-ended items on norms and discomfort. Participants were recruited through snowball sampling, starting from the researchers’ contacts, as well as social media posts and online university communities in Korea. The sample is exploratory rather than representative. Each participant received a gift card worth approximately USD \$3.

We collected 267 responses, excluding 23 during data cleaning (12 from minors, 9 from non-sharers, and 2 lacking consent). The final dataset included 245 valid responses. Participants’ mean age was 34.08 years (SD = 10.37, range 18–64). Age distribution was: teens = 3, 20s = 95, 30s = 80, 40s = 43, 50s = 21, 60s = 3. Gender was 40.8\% male, 54.3\% female, 4.9\% undisclosed. Respondents primarily used ChatGPT (75\%), followed by Claude (11\%), Perplexity (8\%), and Midjourney (6\%), reflecting the Korean market distribution.

Open-ended responses were analyzed using thematic coding. The first author developed an initial codebook, which two independent coders then applied to the full dataset. We split free-text responses into meaning units and applied multi-label coding at the segment level; 20\% of segments were double-coded. Inter-coder reliability was assessed with Cohen’s $\kappa$ = .78, indicating substantial agreement. Discrepancies were resolved through discussion.

\subsection{Interview}
To deepen insights into implicit norms and observer effects (RQ2–RQ3), we conducted semi-structured interviews with 36 participants. We recruited 20 participants from survey respondents who consented to follow-up, selecting purposively for variation in sharing types and relationships. An additional 16 were recruited through referral sampling. Interviews lasted 45-60 minutes and were conducted in Korean via Zoom. Questions addressed the background of account sharing, perceptions of implicit rules, experiences of observing or being observed, and strategies for managing traces and visibility. See the Appendix for the final version of our interview protocol. Participants received USD \$17, plus an additional \$3 for each referral. In total, interviews yielded 45 hours of audio recordings, 530 analytic memos, and 764 transcript segments. Coding followed a bottom-up approach. We first conducted open coding, then refined categories around sharing types, norms, and observer effects. Reliability was ensured through iterative discussion and consensus building; intercoder agreement for the main themes reached Cohen’s $\kappa$ of .79.

As informal coordination patterns emerged during the study, we adjusted the protocol to explore them more thoroughly. Instead of asking about informal coordination directly, we used Malone et al.’s work to gather participants’ stories about their own experiences. Questions were further refined based on participants’ roles within their shared accounts.

Table 1 presents the demographics of the 36 participants, whose ages ranged from 22 to 53. Among the interviewees, 15 were account owners (who paid for and managed the subscription), 19 were members (who shared use without holding the account), and 2 were lendees (who accessed the account through another member’s invitation). Their occupations included homemakers, undergraduate and graduate students, pharmaceutical sales representatives, graphic designers, freelance writers, engineers, developers, photographers, tour guides, product owners, nurses, government employees, and elementary school teachers.

Data analysis employed a systematic, bottom-up coding approach based on field notes and transcripts. The coding process was carried out by the authors themselves, using Excel as the primary tool, without LLM assistance. We first performed open coding collaboratively, grouping relevant responses into emerging themes. As the analysis developed, a coding framework was created around modes of shared use, perceptions of norms, and observer effects related to sharing. Researchers revisited transcripts to refine these themes, discussing and reaching consensus to ensure reliability. See the Appendix for the final version of our interview codebook. To evaluate coding consistency, we calculated the percentage agreement following Graham et al. \cite{Graham2012InterRater} and confirmed intercoder reliability with a Cohen’s $\kappa$ of 0.79, indicating strong agreement.

%% file: tables/table1.tex
\begin{tabular}[\textwidth]{p{1.5cm}p{2.5cm}p{1.5cm}p{1.5cm}}
\toprule
\textbf{Participant (Role)} & \textbf{Platform} & \textbf{Length of Use (months)} & \textbf{\# of Users Recognized} \\
\midrule
P01\_M & ChatGPT & 3  & 3 \\
P02\_O & ChatGPT & 2  & 2 \\
P03\_O & ChatGPT & 5  & Unknown \\
P04\_M & Perplexity & 15 & 2 \\
P05\_M & ChatGPT & 3  & 7 \\
P06\_O & Claude & 3  & 2 \\
P07\_M & ChatGPT & 9  & 2 \\
P08\_O & ChatGPT & 6  & 2 \\
P09\_O & Perplexity & 4  & 3 \\
P10\_O & ChatGPT & 5  & 3 \\
P11\_M & ChatGPT & 30 & 2 \\
P12\_M & ChatGPT & 11 & Unknown (max 6) \\
P13\_M & ChatGPT & 5  & 10 \\
P14\_M & ChatGPT & 6  & 3 \\
P15\_O & Claude & 2  & 2 \\
P16\_O & ChatGPT & 5  & 3 \\
P17\_M & Perplexity & 6  & 2 \\
P18\_L & ChatGPT & 10 & 2 \\
P19\_M & Claude & 6  & 4 \\
P20\_M & ChatGPT & 4  & 6--7 \\
P21\_M & ChatGPT & 7  & 2 \\
P22\_O & ChatGPT & 5  & 2 \\
P23\_M & ChatGPT & 6  & 2 \\
P24\_M & ChatGPT & 8  & 6 \\
P25\_L & ChatGPT & 12 & Unknown (max 7) \\
P26\_O & ChatGPT & 3  & 3 \\
P27\_O & ChatGPT/ GenSpark & 12 / 6 & 2--3 \\
P28\_M & ChatGPT & 6  & 3 \\
P29\_O & ChatGPT & 2  & 2 \\
P30\_O & ChatGPT & 2  & 2 \\
P31\_M & ChatGPT/ Midjourney & 6 / 3 & 6 / 3 \\
P32\_O & ChatGPT & 5  & 2 \\
P33\_M & ChatGPT & 4  & 6 \\
P34\_M & ChatGPT & 12 & 3 \\
P35\_M & ChatGPT/ Midjourney & 5 / 5 & 5 / 2 \\
P36\_O & ChatGPT & 17 & 2 \\
\bottomrule
\end{tabular}

%% file: sections/04-findings.tex
\section{FINDINGS}
\subsection{Descriptive Statistics}
\subsubsection{Sharing Relationships}
Most respondents (92\%, multiple responses permitted) reported sharing accounts with known individuals: family, friends, or coworkers. Among all respondents (n=245), family was the most frequently mentioned partner (45\%), followed by friends (26\%) and coworkers (23\%). Most (89\%) shared with only one group, predominantly family (40\%). By role, 120 respondents (49\%) identified as account owners, and 125 (51\%) as non-owners (members). Owners most often shared with family or friends, while members were more likely to report sharing with coworkers than owners. A small portion of respondents (8\%, 20/245) shared accounts with strangers; of these, 14 (70\%) reported using someone else’s account, all of which were ChatGPT accounts. The number of co-users reported ranged from 2 to 20 (M = 2.91, SD = 1.68), although about 20\% of respondents said they did not know the exact group size.

\subsubsection{Cost Arrangements}
We classified cost arrangements as either casual-sharing (no fee split) or cost-splitting. Among owners (n=120), 64\% engaged in casual-sharing and 36\% in cost-splitting; among members (n=125), the distribution was 53\% and 47\%, respectively. Within casual-sharing, patterns varied significantly by role: 96\% of owners said they paid the full cost, while 94\% of members reported paying nothing. Owner-paid casual-sharing mainly involved family, and, secondly, friends. For non-owners who did not contribute financially,  family and coworkers appeared at similar rates, and in some coworker cases, the subscription was paid by the organization.

\subsubsection{Presence of Norms}
Our analysis of participants' responses revealed three main contents of coordination norms in shared accounts: privacy, boundary setting, and access rules. As shown in Table 2, privacy was the most commonly cited norm (63\%), articulated as avoiding access to others' chat histories and limiting personal disclosures. Notably, the responses extended beyond prohibitions on viewing others' data to an awareness that one's own records might be visible. Hence, statements like “delete sensitive questions yourself” rested on that assumption. Privacy also involved refraining from intruding on others’ ongoing projects, indicating that it served not only as a rule about data access but also as a broader norm for interaction.
The second category, boundary setting (24\%), involved norms that addressed mixing traces when multiple users shared a single account. Examples included “do not delete others’ work at will” or “create new chats only within one’s own folder.” Some participants used ChatGPT’s project feature, colloquially called “folders,” to separate and manage individual conversations.
Finally, access rules (13\%) appeared as efforts to prevent disruptions caused by system limits, such as monthly query caps or concurrent sessions. For example, in models like GPT-4.5 that enforce monthly usage limits, some groups negotiated clear quotas to share usage fairly. These norms were sometimes openly discussed among co-users, but often they were applied implicitly on an individual basis. Responses like “just my own idea” or “that’s just how I see it” indicate that not all norms were negotiated collectively or shared widely.

\begin{table}[]
    \centering
    \caption{Contents of coordination norms identified in shared LLM accounts. Responses were coded into three main categories: privacy, boundary setting, and access. The table shows representative example responses. Unit = coded segments; primary code reported per segment. Totals: S=91 segments, P=245 participants. Counts and shares: Privacy 57 (63\%), Boundary setting 22 (24\%), Access 12 (13\%). “\% seg.” uses S; “\% part.” counts unique respondents per code. 20\% double-coded (Cohen’s $\kappa$=0.78).}
    \label{tab:table2}

\input{tables/table2}
\end{table}

\subsection{Four Types of Account Sharing}
Our analysis reveals four distinct types of LLM account sharing based on two factors: whether the account owner actively uses the service and whether costs are shared among members. The categories include: owned-casual, ownerless-casual, owned-split, and ownerless-split. The first factor pertains to whether the account owner—the individual who pays for the subscription—also actively uses the account. The second factor relates to whether group members split the subscription fee. This typology provides a structural framework for understanding how motivations, trust relationships, and coordination strategies vary across different forms of account sharing.

\subsubsection{Owned-casual}
In owned-casual sharing, account owners allow others to use their service primarily for altruistic reasons. In this setup, the owner did not ask for payment from friends or family members. Eight participants fell into this category. Group sizes ranged from two to four members, sometimes including entire families. Usually, all members were composed of individuals known to the account owner.

In these cases, the account owner served as the practical manager of settings, payment, and security. While they were under no obligation to share, they often did so out of altruism or a sense of relational responsibility. One participant explained: \textit{"I found it so useful that I recommended it to a friend. When they mentioned the \$20 monthly fee was burdensome, I offered to share mine."} (P06\_O). Another remarked: \textit{“It was just sharing within the family—I simply entered the password into my husband’s phone so he could log in.” }(P12\_M).

In owned-casual arrangements, both parties typically view sharing as a temporary favor. Sharing norms emerged implicitly between account owners and members. However, since the owner bore the full cost, they sometimes set explicit usage boundaries that others were expected to follow. In addition, chat outcomes frequently extended into offline conversations—such as over meals or in meetings—and the communication of rules and announcements was often handled face-to-face rather than through external channels.

\subsubsection{Ownerless-casual}
Ownerless-casual sharing occurs when the account holder does not use the service themselves. This often occurred when a company or organization holds the subscription and provides access to members at no cost for work-related reasons. Interviews revealed that many organizations try to save money by sharing a single account instead of purchasing individual licenses. As one participant said:\textit{ “Here, it feels like we spend millions of won every month on subscriptions, almost like paying an electricity bill. It’s one of the most critical parts of the company’s infrastructure.”} (P35\_M)

Group sizes ranged from small teams of five to departments of thirty employees across a department or the entire company. Because the account was organizational from the start, there was no strong sense of individual ownership. Account management and monitoring were usually loose or minimal. A team leader explained: \textit{“As the team leader, I wouldn’t say I really manage the account—I just keep an eye on it. But when it comes to how people actually use it, that’s not something I can control.”} (P05\_M).

Information about account use was usually shared through workplace communication channels such as Slack, group messengers, or internal email lists. Formal rule documents existed but were largely symbolic, with little practical influence on daily use. As another participant described:\textit{ “At our company, there are separate systems for task and document management that everyone logs into. The account information is stored there, and activities across different teams and departments are all recorded and traceable. Because history is also linked to what we do in ChatGPT, we emphasize asking specific and accurate questions. But in practice, the level of prompts people write varies widely.”} (P13\_M).

Norms were a mix of written rules and practical habits. Announcements usually outlined principles like banning third-party access and discouraging the disclosure of sensitive information, but these enforcement measures relied mostly on team agreements. Managers seldom enforced strict rules, instead watching usage and stepping in as moderators only when conflicts or overload occurred.

\subsubsection{Owned-split}
Owned-split sharing describes situations where the account owner regularly uses the service and shares subscription costs with others. Participants in this group often shared accounts to reduce the relatively high fees compared to the perceived benefit. As one participant mentioned:\textit{ “I saw a post on a university community board asking, ‘Anyone want to share ChatGPT?’ so I left a comment and joined. The main appeal was how much cheaper it became. I thought sharing with multiple people might cause traffic issues, but in practice, it felt no different from using it alone, which was surprisingly fine.”} (P33\_M). Another participant reflected on continuing the sharing arrangement from previous subscription methods: \textit{"Initially we discussed splitting costs due to the high fee (…) and then we just kept using it, almost like sharing a Netflix account.”} (P17\_M).

Some participants viewed premium plans as having capacities that are excessive for individuals but valuable when shared. One explained: \textit{“ChatGPT Pro is great, but paying for it solo feels wasteful. I can't use all that capacity by myself … I rarely reach that limit by myself, but when sharing, those higher capacities become useful. It feels less like overpaying and more like supplementing each other’s needs.” }(P27\_O)

The account owner usually managed payments and credentials, while sharing was most common among peers like friends or colleagues. However, some participants preferred sharing with less familiar people rather than close friends, citing privacy concerns. One participant remarked: \textit{“What bothers me most is the idea that my private life might be exposed. It feels a bit easier knowing I’m not using it directly with my brother but rather with one of his work acquaintances I don’t really know. That made me more comfortable about sharing the account.” }(P29\_O).

Unlike owned-casual, owned-split members often maintained dedicated outside communication channels to handle payments and credentials. As one participant explained: \textit{“When the account owner posts in the group chat, ‘This month’s cost is about \$4,’ I transfer the money and mark the message to show it’s done.” }(P33\_M). Norms were sometimes made explicit through these channels, but many participants felt that spelling out the rules in detail felt uncomfortable. As a result, only minimal guidelines were set, with the rest managed implicitly.

\subsubsection{Ownerless-split}
Ownerless-split sharing involves third-party platforms managing subscriptions for user groups. Instead, a company or intermediary platform handles subscription, account management, and billing while collecting fees from participants. In this setup, participants were usually strangers or nearly anonymous to each other. One participant explained: \textit{“There is a platform called Gamsgo, a global subscription-sharing service. On this site, people can form groups to split subscription accounts—not only for ChatGPT but also for services like Netflix. Sometimes the groups include Koreans, and other times international users as well. When I make the payment, the platform creates an account based on the applicants at that point and sends the login details to each person’s email.”} (P31\_M)

Most participants had previous experience sharing other subscription services to save money and knew where to find potential co-users. Services such as Gamsgo exemplify a subscription-sharing platform that transforms conventional individual subscription models into multi-user environments, providing an economical alternative for users deterred by official pricing. Once registered and subscribed to a specific service, users are automatically matched into groups; the platform then periodically refreshes and redistributes login credentials to subscribers. Although options where the owned-split access were also available, some preferred the anonymity of sharing with strangers rather than sharing closely with friends. As one participant recalled:\textit{ “One friend suggested, ‘Since we trust each other, the four of us should share an account.’ But I actually felt it might be better with strangers. With close friends, I became more self-conscious, almost shy. Sharing with people I don’t know felt more anonymous, and in that sense more comfortable.” }(P31\_M). The group size was determined by the platform’s default settings, usually ranging from three to six members. Since members communicated through an intermediary platform, there were no dedicated channels for direct contact. Norms developed almost entirely implicitly, with participants mainly aligning their behavior with existing usage patterns. For example, newcomers often copied folder structures already created by others or saw their own habits gradually become normalized as shared conventions. Basic attitudes or courtesies—such as “do not look at others’ records” or “avoid entering unnecessary sensitive information”—also became norms.

Thus, norms in account sharing varied significantly depending on whether the owner actively participated and if costs were shared. When the owner both paid for and used the account, authority and responsibility were prioritized, and norms were often explicitly established by the owner or minimally defined boundaries were set (owned-casual, owned-split). In contrast, when the owner did not use the account directly, management was more relaxed, and norms tended to depend more on collective practice or implicit agreement (ownerless-casual, ownerless-split).

Cost-sharing also acted as a way to promote fairness and responsibility in managing accounts: when fees were split, members engaged in more active negotiations and kept dedicated communication channels. Norms, therefore, were not fixed rules but practices that developed and adjusted relationally, influenced by the conditions of owner participation and cost structure. In the next section, we explore which specific norms emerged under these conditions and how they were created and maintained through both explicit and implicit methods.

\subsection{Types and Formation of Norms}
As a form of appropriation, account sharing involves users extending services beyond their intended design and establishing their own norms in the process. In our study, sharers formed norms around boundary setting, access rules, and privacy. Earlier, we classified account sharing into four types, based on whether the owner actively uses the account and whether costs are shared; these two axes critically shaped how norms were formed. At the same time, prior work has shown that implicit norms are inherently fragile due to their lack of documentation and the social costs of enforcement; in our context, this fragility interacted with the visibility of shared traces, giving rise to additional tensions.

\subsubsection{Norm Types: Explicit vs. Implicit}
\paragraph{Explicit Norms}
Explicit norms emerge when members openly discuss and state rules for managing shared accounts. The two structural axes—whether costs are shared and whether an active owner is present—shaped both the degree of explicitness and who articulated the rules. We found two types: co-negotiated norms and owner-imposed norms. Co-negotiated norms appeared mainly in casual arrangements with a single payer, where members already knew each other and relationships were relatively equal. In these cases, users often agreed on rules beforehand to ensure fairness. On the other hand, in \textbf{owned-casual} settings we occasionally saw the payer/owner put forward unilateral rules. As one owner put it: \textit{“It’s my account and I’m the one paying for it. Since the cost isn’t being shared, I feel I should keep full control—for example, the authority to delete anything.”} (P02\_O). This shows how owners sometimes set rules to reinforce their sense of ownership as the payer. However, many participants felt uncomfortable making rules explicit for fear of relational friction. One participant explained their reluctance to establish explicit rules: \textit{"Bringing that up could make everyone uncomfortable [...] So even though something bothers me, I act like it doesn't and let it slide" }(P32\_O). As a result, explicit rules often settled in partial, incremental, or auxiliary forms; in many cases, norms were instead set and elaborated implicitly.

\paragraph{Implicit Norms}
Implicit norms emerged either through reading collective practices or through individually established routines. They appeared across all sharing types but were most pronounced in \textbf{ownerless-split} settings, where all members shared costs and no owner was present. Transmission and learning occurred through a variety of indirect channels—off-platform chats, offline conversations, and especially observation and mimicry. Notably, even when external channels were available, participants were often reluctant to articulate norms explicitly; in cases where no such channels existed, snooping frequently became the means through which norms were learned and enacted. For example, one participant remembered: \textit{"Without direct communication channels, I observed that five others had created project folders and followed their example."} (P31\_H).

Individually defined norms were also set without prior coordination or evolved over time, and they sometimes extended into expectations about others. Sometimes account owners believed others would follow their own habits: \textit{“In my case, I never really entered personal concerns or private conversations into ChatGPT, and I just assumed others wouldn’t either.” }(P10\_O). In other cases, participants based their behavior on notions of politeness:\textit{ “I feel it’s only polite not to leave my traces in something another person has used. If someone else’s marks showed up in my own record, even among people I’m sharing with, it would feel a bit unpleasant. And I assume others feel the same way.” }(P29\_O). Still, some formed personal norms through direct experiences within the account, shaping practices based on past incidents.

\subsubsection{Fragility of Norms}
Users frequently violated norms, with implicit norms proving especially vulnerable. Because they leave few traces of agreement or responsibility, their content is hard to see; oral or off-platform agreements (e.g., via messengers) also evaporate over time. Key terms such as “sensitive information” invite divergent interpretations, and when no communication channel exists, violations are hard to verify and accountability is delayed. As a result, members infer rules from their own experiences and internalize different versions, reproducing gaps in norm awareness even within the same account. This was evident in interviews: a married pair sharing one account (P22\_O / P23\_M) gave opposing answers—one saying “\textit{you can look at others’ chats,”} the other insisting that\textit{ “privacy should be respected, even between spouses.”}

Violations clustered around privacy. While some participants reported adding extra devices or lending credentials to outsiders, \textit{snooping} stood out because it was hard to detect: viewing others’ conversations often left no logs or notifications. Under such low-visibility conditions, people opened others’ chats despite agreeing “not to look.” In fact, 30 of 36 interviewees admitted doing so. As one participant put it: \textit{“When you look into other people’s chats, it feels quite uncomfortable, but—how should I put it—it was almost voyeuristic, like peeking at how others live. You can’t help but notice how people use it, since those one-line titles keep appearing.” }(P25\_L)

Privacy also entails an enforcement paradox: verifying compliance requires visibility into content or metadata, yet that verification itself risks infringing privacy. Moreover, the benefits of snooping are personal and immediate, whereas the \textit{costs}—erosion of trust or relationship strain—are structural and delayed, leaving the rule especially exposed. Layered together, these conditions made privacy norms crack first.

Users were acutely aware of these weak points. Anticipating that their traces might be seen, they pre-emptively adjusted what they wrote—refining tone and omitting sensitive details—thereby strengthening self-censorship. At the same time, they rationalized peeking as a way to check others’ compliance. Potential observability thus amplified both self-censorship and mutual monitoring, further thinning the privacy norm. This dynamic intensified where costs were shared (heightening demands for fairness checks) and where an active owner existed (normalizing management/audit exceptions). We analyze these interactions through the lens of the observer effect.

\subsection{Observer Effect: as-Actor and as-Target}

\begin{figure*}
    \centering
    \includegraphics[width=0.8\linewidth]{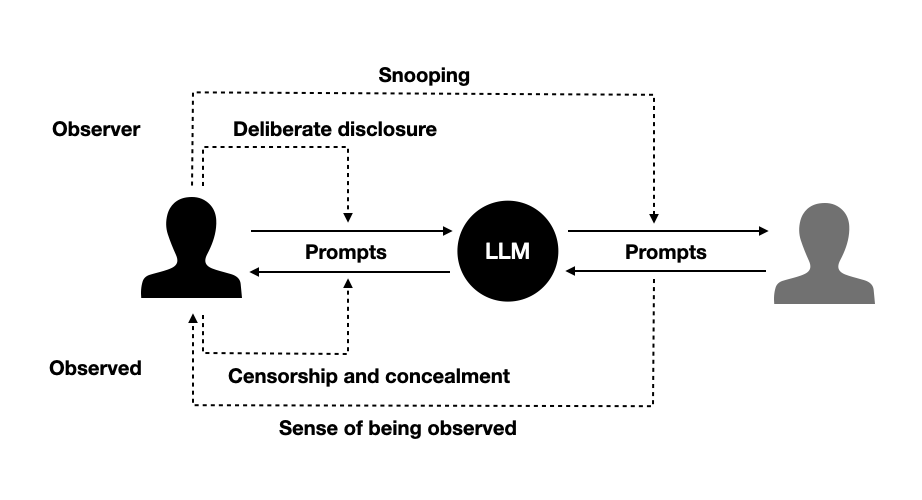}
    \caption{Illustration of the observer effect in shared LLM use. The figure shows how users act as both observers and targets: as observers, they may snoop or deliberately disclose prompts; as targets, they experience a sense of being observed, leading to censorship and concealment. These dynamics structure how prompts are produced and interpreted within shared accounts.}
    \label{fig:figure}
\end{figure*}

The observer effect (see Figure 1) refers to changes in user behavior caused by the presence and traces of others, as well as the possibility of being watched. It includes both actively viewing someone else’s conversation and the awareness that one’s own interactions might be observed, prompting users to alter their behavior. In our context, this mainly involves privacy boundaries: users both peek at others’ conversations and expect to be watched, which influences how they set and enforce these boundaries.

The observer effect manifests in two interconnected ways. First, observer-as-actor describes how users read and interpret others’ traces and then adjust their own actions accordingly—often by peeking to check fairness, learn, or satisfy curiosity—facilitated by always-visible lists of recent chats, folders, and token indicators that can be accessed with a few clicks. Second, observer-as-target illustrates how users internalize the possibility of being observed, leading to self-censorship and self-regulation—such as softening or deleting sensitive queries, or controlling what and when they post. These effects are not mutually exclusive; they occur together and reinforce each other, especially in shared environments with partial visibility.

\subsubsection{Observer-as-Actor Effect}

The observer-as-actor effect refers to situations where users detect and interpret others’ traces in a shared interface, and adjust or intervene in their own usage based on those interpretations. This effect arises in environments where chat histories are naturally exposed: users go beyond the original scope of sharing, interpreting, and appropriating others’ actions and traces to regulate their own behavior. In doing so, they move from being mere co-users of the service to acting as interpreters of others’ use. This includes not only direct inspection of conversations but also inference from weaker cues such as sidebar titles, folder structures, or recent activity indicators.

Participants described both intentional and unintentional snooping. Intentional snooping mainly happened for four reasons. First, curiosity denotes a state of mind, rather than having a specific interest. Curiosity led users to explore the types of questions others asked and in what context, transforming account sharing into a means to peek into others’ knowledge-seeking behaviors and interests. As one participant recalled: \textit{“Whenever my mom had a gathering—whether with family or friends—at first she kept generating Ghibli-style images using photos of the people there. She didn’t seem satisfied, so she kept trying with different photos, even old ones. I remember laughing a lot as I watched those threads.”} (P29\_O). 
Second, some participants engaged in account sharing not out of general curiosity, but to follow up on a specific person or issue. This involved checking how often or in what ways another user interacted with the service, using it to infer their habits or activity levels.\textit{ “There’s quite a big age gap between us, and in a way I’m like a half-caretaker for my younger sibling. So yeah, I do check in on how she’s using it. If they’re not using it right, I feel like I should step in and teach her.”} (P26\_O) Third, checking access status focused on the meta-condition of the account, which related to performance concerns: since simultaneous use could slow the system, some users looked to see if others were currently active before logging in themselves. These deliberate actions ranged from casual curiosity to strategic efforts to improve efficiency.\textit{ “I sort of have a routine of checking the member’s project folder, maybe once a day. There’s this thing — I get a little worried that I might’ve used it too much and the other person couldn’t use it. Technically, it’s my account, but I do worry if it causes any trouble.” }(P32\_O) Lastly, some participants described intentional viewing as a way for learning. They actively observed phrasing and structure of co-users to improve their own use. One participant recalled:\textit{ “One of the people I share with is a mathematician, and he'd write that he was a university math student interested in matrices. After seeing that, I also started by writing, ‘I’m a UX/UI major named \_\_\_,’ and then continued with my prompt. That way, personalization could work for me as well.”} (P28\_M). In these
cases, what might otherwise be snooping functioned as informal peer learning.

In contrast, unintentional snooping happened through accidental clicks or page transitions, which revealed only small parts of someone else's conversation. As one participant explained: \textit{“I clicked on it, almost by accident, and only one or two of the most recent chats popped up. That was it."}(P14\_M).

Users who viewed others’ conversations described mixed emotional reactions, including guilt, discomfort, or resentment. For example, one participant recalled the unease of seeing her child’s private conversations:\textit{ “After my child was scolded, he had written down his inner feelings. (…) Seeing those thoughts made me uneasy—as if it would have been better not to have read them at all.”} (P09\_O). By contrast, some viewed snooping as more helpful in understanding the situations or emotional states of close friends and families. One participant (P34\_M) learned that her brother had gastroenteritis through shared conversations and considered the information quite helpful.

Participants also described responding directly to the authors of conversations they observed. For example, one participant (P01\_M) mentioned that his colleagues once joked about—and even mimicked-the “fortune-telling” request they saw in ChatGPT. Observation sometimes went beyond passive reactions to active changes in practice. Users took the information they encountered as learning material and applied it to their own work. One participant explained: \textit{“I read through some writings by an astronomer, which were in English. I translated them into Korean and then, as I mentioned earlier about prototyping a design, I incorporated those texts into my scenario—asking ChatGPT to adapt them in a more engaging way.” }(P28\_M). Others reported learning new prompting techniques by watching how co-users wrote. One participant reflected: \textit{“Before, I'd just say 'shorten the text'. Then I saw someone specify things like 'make this 1,000 characters' or 'write 500 characters.' After that, I started saying 'reduce it by 30\%' or 'by 50\%' instead of just 'shorten it.' People always say your instructions (prompts) should be specific, but I hadn't understood what that meant. Watching how the other person wrote made it click-this is what it looks like." }(P32\_O)

These experiences demonstrate why even after internalizing the norms, users kept peeking others’ conversations. In environments where rule-breaking rarely led to punishment, the power of norms diminished. Users started to believe they “wouldn’t get caught,” and some even began to see snooping as normal or acceptable. As a result, snooping became widespread on shared accounts, increasing the gap between what people claimed was right and what they actually did.

\subsubsection{Observer-as-Target Effect}

The observer-as-target effect involves behavioral changes arising from users' awareness of potential observation. In shared-account settings, seeing others' traces often makes users conscious of their own visibility. This realization shifted their view from mere users of the service to potential subjects of evaluation and interpretation, creating a psychological burden. As one participant admitted: \textit{“I do worry about security—like, what if someone sees what I’ve been asking?”} (P31\_M).

The possibility of being observed influenced the topics users asked, how they phrased messages, and the level of detail in their instructions. Before writing, many wondered, "How will this look if someone else reads it?” Two suppression strategies emerged: censorship and concealment. Censorship occurred before writing, with participants changing their tone or avoiding sensitive subjects. One participant explained, \textit{“Because I’m sharing the account with others, I feel less free in how I use it. Knowing that co-users could see every detail of my wording, I find myself trying to use more formal, polite language" }(P04\_M). Concealment happened after writing, through deleting or editing chat histories or using temporary chat modes. Some even created separate free accounts solely to ask a few private conversations.

These suppression practices were evident across all four sharing types. In casual setups, users were especially cautious about revealing trivial details or “Too-Much-Information” to people they knew. One participant noted: \textit{“I saw colleagues using it for things like sharing personal worries or even playful questions like asking GPT about their MBTI type. But since I’m sharing with siblings or people I know, I purposely avoid that. Even if I weren’t sharing, I probably still wouldn’t—I’d hold back.” }(P10\_O). In split arrangements, concerns focused more on personal data leakage:\textit{ “For fun, I might want to ask it to read my fortune based on my birthdate. Revealing my job might be fine, but in a shared account, I hesitate with anything more personal. I think about asking, then stop myself.” }(P14\_M).

Rather than prompting self-censorship, some participants leveraged it as an opportunity, deliberately leaving messages to shape how others perceived them. This strategic self-disclosure was driven by a desire to express opinions, show knowledge, or establish their presence. One participant said: \textit{“I think part of me actually wants others to see it. If someone else looks at it, maybe it'll rub off in a good way.”} (P03\_O). These performances included carefully crafted remarks designed with an imagined audience in mind. Another participant explicitly took English-learning exercises as a kind of presentation:\textit{ “For example, I might write something like, ‘Can you practice English conversation with me for 30 minutes?’ Knowing that, I sometimes tweak the wording so my English comes across a bit better.” }(P05\_M) 

These examples show how visibility strategies gained significance within relationships, turning the shared account into a space for mutual understanding and relationship-building. Interestingly, these disclosure strategies were found across all four types of accounts. In casual sharing, disclosure largely focused on relationships, aiming to affirm presence or strengthen ties. In split sharing, it served a more practical role: providing prompts or Q\&A processes for others to reference. As one participant explained: \textit{“By looking at other people’s Q\&A, I can get information without having to ask the question myself. That’s what I see as the benefit of sharing. If the topic is similar, I can even reuse the prompt with slight modifications. I genuinely hope that everyone sharing the account benefits from it.”} (P12\_M).

\subsubsection{The Recursive Cycle of Acting and Being Observed}

The two aspects of the observer effect reinforced each other in a cycle, and users alternated between observer and observed. Reading others’ traces increased the sense that “others could be looking at me in the same way,” which intensified the observer-as-target effect. In this way, users shifted between the roles of seeing and being seen, making the shared account a unique space of mutual surveillance. As one participant described: \textit{“Every time I look at someone else’s record, it feels like a mirror—I can’t help but think they might be looking back at me in the same way. In that space, you are both the watcher and the watched, reflections bouncing back and forth without end.” }(P12\_M)

This cycle brought lasting changes to interaction patterns and chat-history management. On one hand, the shared environment became a stage for self-presentation, where users left chats not just for utility but also to signal diligence or expertise. Everyday conversations were transformed into acts of self-management and image building, turning shared accounts into sites of social performance. Sometimes, these disclosures even carried over into direct interpersonal exchanges. One participant recalled:\textit{ “At the time we weren't really seeing each other- we'd had a fight and were only talking over chat. After an argument, I once asked the LLM for a second opinion, typing something like, ‘In this situation, isn’t she the one at fault?’ But my girlfriend actually asked me directly, ‘I don’t see it that way—why would you put it like that?’”} (P27\_O). Thus, audience-aware utterances functioned as strategies of facilitation.

These dynamics created opposing strategies for managing chat histories, facilitating self-presentation on one hand and censorship or suppression on the other. Being constantly under others’ gaze was likened to maintaining a particular persona, which many found uncomfortable. Some carefully edited their wording or deleted and revised records to reduce exposure: \textit{“I tend to delete everything—so people won’t form unnecessary impressions. If they see it, some might think ‘she only works,’ while others might think ‘she doesn’t even know this.’ So I just erase it all.”} (P05\_M). For some, the psychological burden was so strong that it pushed them out of the shared space. They set independent free accounts for personal use or even considered leaving the server altogether: \textit{“I don’t like leaving logs, so I rarely use the shared account for personal things. I mainly rely on a free personal account instead.” }(P05\_M)

These contrasting outcomes highlight more than just a gap between norms and behavior. In shared environments, users simultaneously experience positive opportunities—such as learning, inspiration, and relationship-building—while also confronting privacy concerns, anxiety, and the burden of constant self-management. This cyclical pattern under-scores important design implications: how to set boundaries, establish norms, and support observation in reciprocal, selective, and time-limited ways within shared LLM settings. These issues pave the way for the alternative design directions we will discuss next.

%% file: tables/table2.tex
\begin{tabular}{p{1.9cm} p{4.6cm} p{1.0cm}}
\toprule
\textbf{Coordination Norms} & \textbf{Example Responses} & \textbf{N (\%)} \\
\midrule

Privacy &
\textbullet~Avoid viewing others' conversations \newline
\textbullet~Keep conversations private; do not read others' chats \newline
\textbullet~Do not look at other people’s folders (Just my own idea) \newline
\textbullet~Do not pry; that's just how I see it \newline
\textbullet~Do not ask overly private questions \newline
\textbullet~If you ask a sensitive question, delete it yourself 
& 57 (63\%) \\
\midrule

Boundary Setting &
\textbullet~As long as you don't touch the other person's project, there's no problem \newline
\textbullet~It's okay to look, but don't interfere \newline
\textbullet~Do not delete others' work without permission \newline
\textbullet~Do not erase the session history without checking with others \newline
\textbullet~Ask questions only in your own session/folder \newline
\textbullet~Keep session files organized  \newline
\textbullet~Create new chats only in your own folder
& 22 (24\%) \\
\midrule

Access &
\textbullet~No simultaneous use of the account \newline
\textbullet~Use it reasonably; if you don't like the rule, get your own paid subscription \newline
\textbullet~Do not waste tokens recklessly
& 12 (13\%) \\
\midrule

\textbf{Total} & & \textbf{91 (100\%)} \\
\bottomrule
\end{tabular}

%% file: sections/05-discussion.tex
\section{DISCUSSION}

We structure the discussion around our research questions. We first revisit how the four sharing types (RQ1) organize motivations and coordination conditions. We then discuss how implicit norms emerge and fracture across these types (RQ2). Finally, we explain how persistent conversational traces amplify observer effects, shaping both snooping and self-censorship (RQ3).

Our study of LLM sharing reveals that interactions leave behind cognitive traces. Conversation histories show how problems are tackled and how people express themselves emotionally and socially. In shared LLM accounts, where shared users have separate threads for conversations between themselves and the agent, what becomes perceptible are those \textit{cognitive traces}. Important aspects of these cognitive traces include the contextual information within the questions, the narrowing of the prompt's scope, the affective cues such as expressions of gratitude or insults directed towards the agent, and strategies adopted for achieving a more satisfactory answer. These traces capture the process of prompting, which makes problem-solving processes reusable \cite{catledge1995characterizing, Ericsson1993, goyal2016effects, kery2018story, knuth1984literate, maclean2020questions, pirolli1999information, ragan2015characterizing}. Since cognitive traces are rich in meaning and can be considered closely related to thinking, they can serve as resources for learning and coordination. However, they can simultaneously put users at risk of misinterpretation in ways that behavioral logs like clicks or page views do not \cite{dabbish2012social, mamykina2011design}. In this study, we recognize that this multi-user application is a new form of communication and learning, and therefore a form of "cognitive sharing".

Overall, our findings highlight three main contributions to understanding LLM account sharing. First, we show how two key axes—whether the owner actively uses the account and/or whether costs are split—shape not only the norms that develop but also users’ willingness and ability to coordinate. (RQ1) Within these two axes, four common setup types exist, each characterized by distinct reasons for sharing, varying relationships between members, and unique ways of gathering to share an account. While owner participation places responsibility for daily management on the owner, cost-splitting gives members leverage to negotiate fairness. Second, we explain how implicit norms around reasoning traces create an observer effect tension: users can access others’ traces through snooping or borrowing, even as they hide or delete their own traces to avoid judgment. (RQ2) In this study, we recognize that the main way norms are maintained is not just through the interface but also within the cognitive traces. Third, we translate these insights into design suggestions—negotiation tools, social transparency features, and graded visibility controls—that adapt single-user LLM interfaces for multi-user settings while still considering user flexibility. (RQ3)

These contributions extend classic CSCW concepts such as awareness \cite{Dourish1992Awareness}, social translucence \cite{Erickson2000SocialTranslucence}, and articulation work \cite{Schmidt1992}. Prior research emphasized the visibility of activities as the basis for coordination and accountability. Alongside the familiar issue of ‘can’t see’—missing or ambiguous cues—we identify cases of ‘won’t move,’ where traces are visible but potential observers choose not to act because of social costs, risks, or normative uncertainty. Visibility also becomes asymmetric. We differentiate between observer-as-actor, who interprets others’ traces, and observer-as-target, who stylizes or suppresses their own traces, anticipating being seen. Many users switch between these roles of the observer and being observed. These asymmetries highlight that our approach isn’t just redefining existing perspectives but posing a new examination of sharing thought processes as a unique coordination surface. 

Finally, our findings speak to platform-level decisions about whether to discourage or accommodate account sharing. Across subscription ecosystems, some providers have moved toward locking down shared credentials, as in Netflix’s move to restrict multi-household accounts which frames sharing primarily as revenue loss and security risk. However, unlike streaming services like Netflix, some participants reported learning/inspiration as a secondary outcome once sharing was already in place where non-paying members experiment with the service through someone else’s subscription. Simply closing off sharing may not remove these practices; instead, it may push them into more precarious forms, such as third-party group-buy platforms and informal credential-sharing arrangements, where privacy, accountability, and support are weaker. Our findings do not imply that platforms should embrace account sharing as an economic strategy. Platforms may still have valid incentives to enforce single-user models for revenue, security, and policy compliance. However, from a user-centric perspective, sharing remains prevalent and often migrates to informal or third‑party workarounds when unsupported. This suggests a design opportunity: if platforms decide to accommodate some shared-use scenarios, doing so through safer, transparent configurations may reduce privacy and coordination risks observed in our data.

When it comes to LLM account sharing, two levels of appropriation emerge. First, users repurpose individual accounts through credential sharing and setting norms, which are patterns widely documented in domains like streaming services, workplace systems, and household technologies. However, we also observed a second appropriation unique to LLMs: once people begin sharing, they realize that what is being shared is not simply the account itself but also the traces they leave within the account. These then become new objects of sharing as users anticipate how these traces may be assessed by others. It is here that the observer effect starts to shape their behavior. Consequently, they must continuously manage how their cognitive traces are exposed and judged. In this case, appropriation does not only involve the ability to share the account, but also involves the management of the social gaze of those cognitive traces. This second layer of appropriation is what differentiates itself from other forms of account sharing like Netflix, revealing unmet user needs and system limitations that are specific to LLM environments.

At this level, the observer effect creates new challenges, highlighting systematically unsupported social tasks that could facilitate coordination between account sharing. To date, current LLM interfaces only support single-user accounts. Drawing on Gaver et al. \cite{Gaver2003Ambiguity}, who see ambiguity as a design resource, and Tchounikine \cite{Tchounikine2017DesigningForAppropriation}, who emphasizes meta-structures for appropriation, we outline three design strategies that preserve interpretive flexibility while providing safe structures for distribution. Accordingly, Section 5.1 discusses account sharing as a space for negotiation, Section 5.2 explores how social translucence can make implicit norms visible and sustainable, and Section 5.3 presents graded visibility controls that protect intellectual vulnerability and may enable learning while mitigating anxiety.

\subsection{Designing for Appropriation as a Space of Negotiation}
Across sharing types identified in our study (owned-casual, ownerless-casual, owned-split, ownerless-split), participants coordinated through three relationship patterns: close ties (offline), semi-anonymous groups (external channels), and anonymous groups (no direct channels). Firstly, LLM sharing between close relationships often extends to offline discussions. Secondly, semi-anonymous relationships, such as work colleagues or acquaintances, often see coordination through external open channels like Slack. Finally, in anonymous relationships, no formal channels of communication emerge. P26\_O noted, \textit{“These days I use ChatGPT almost every day, and I worry that if I use it too much, my younger sibling might not get their turn.”} However, when agreements and rules depended on external messengers or verbal conversations, the probability of these norms being forgotten increased, which led gaps in shared understanding and re-coordination. \textit{"I told (them) not to look, but I know they do. It's pointless to tell them what to do. I won't even notice it anyways."} (P26\_O) As such, our study identifies an opportunity to build supportive communication systems in the LLM itself. These findings make it clear that account sharing is not just a matter of convenience but a form of appropriation, that is, an active way users reshape systems to adapt to their social needs.

In practice, appropriation involves various motivations and interests, necessitating ongoing negotiation among users. This perspective aligns with the idea of "design-in-use", which identifies that technologies are continuously reinterpreted and modified after deployment \cite{Henderson1991DesignAtWork, Salovaara2006StudyingAppropriation}. This echoes prior research on the everyday appropriation of technology in the domestic sphere \cite{FrohlichKraut2003SocialContext, BrushInkpen2007YoursMineOurs}. However, previous design solutions often concentrated on standardized options like multi-profile accounts, which risk oversimplifying the complexity of the system usage. 

Across the four sharing types, participants struggled to align expectations because coordination relied on ephemeral chats and/or offline conversations. Systems could address this by providing dedicated spaces for recording "house rules" (e.g., deleting sensitive questions, avoiding simultaneous logins) where update notifications are provided to all members. Additionally, contextual annotations on folders or conversations would allow users to signal intentions (e.g., "please don't delete this thread, ongoing project") without indirect coordination elsewhere. While these features could transform fragile off-platform negotiations into sustainable practices, design must avoid over-formalizing arrangements, such as the previously discussed construction of multi-profile accounts. Prior work shows that implicit norms derive their value from adaptability, where users negotiate boundaries through situational practices rather than fixed protocols \cite{Chandrasekharan2018RedditNormViolations, Fiesler2019FanNorms, Rashidi2020SanctioningNorms}. In this case, effective design allows for more visible coordination while preserving flexibility, enabling ongoing renegotiation.

Concretely, our findings suggest that different actors may benefit from different default powers to propose and ratify "house rules". In owner-paid casual accounts, the owner is the natural initiator of rules, but we propose that the interface should invite co-users to acknowledge or contest them (e.g., a rule can only be implemented when everyone has engaged with it and then optionally confirmed it). In cost-splitting groups, the construction of regulations should be collective, where any member can propose a new rule. It is only then marked as “active” once a majority of members have confirmed it. In organizational accounts, administrators define a baseline policy, while teams can then go on to define local rules within specific workspaces or projects. Across these settings, the system can mediate rule creation. For example, one conflict that might arise is if two users defined one thread separately as “ephemeral” and “archival”, prompting the group to revise the regulation rather than allowing contradictions to persist.

Ultimately, the goal of design should not be to suppress appropriation, but to create spaces of negotiation where it can happen safely and effectively. This redefines account sharing beyond the limitations of the single-user model, grounding it in real-world usage. The benefits of this design interpretation emerge in three main areas. First, for users, clear tools for setting and rule agreements can decrease unnecessary conflicts and build trust. Second, for platform policy makers who dictate regulatory and legal frameworks; offering these negotiation spaces can help to redefine what was previously considered a “violation” of regulation into a more transparent form of usage. In this case, they are able to more actively and efficiently provide policy changes based on the way that users themselves interact with the system. Finally, for the designers of the platform, this new perspective considers account sharing not as a problem that requires a solution, but rather an opportunity to advance the utility for its users. This reframing highlights an opportunity to better align LLM platforms with the social and emotional realities of shared use, while still enabling incidental learning and collaboration when users choose to make traces visible. 

\subsection{Making Implicit Norms Visible and Sustainable}
Participants in the study reported that the most prevalent issue with shared LLM usage was the fair division of usage between the users. That is, some users consume a disproportionate share of the tokens. In order to prevent this, users generated implicit norms in the account, which, despite the attempt, were often found to be fragile and incomplete.

This gave rise to three key vulnerabilities. First, the emergence of conflict avoidance: \textit{"I thought about changing the password, but didn't bring it up because I was afraid it might cause an argument."} (P32\_O). Second, the construction of vague boundaries: \textit{"When my younger sibling is using it, I try not to log in, but there isn't really a rule—it's just about reading the situation"} (P10\_O. Third, the surfacing of attribution anxiety:\textit{ "With a shared account, you can't tell who did what, and checking feels like surveillance" }(P03\_M). These examples illustrate how implicit norms remain incomplete and fragile.

Previous research documents similar vulnerabilities where certain changes in both internal and external context can cause sudden and unpredictable norm deterioration \cite{Cheng2017Troll, Dym2020NormVulnerability}. Studies on online communities show this is most prevalent through factors such as an increase in member numbers, system setting changes, or costs \cite{Chandrasekharan2018RedditNormViolations, Rashidi2020SanctioningNorms}. 

As such, rather than imposing rigid rules, design efforts should support a natural development of user norms. This could be done through the following translucency features: (1) presence indicators which identify the activity of certain users within the system to prevent simultaneous usage; (2) activity summaries which provide a brief overview of chats to allow users to collectively track patterns within the LLM; and (3) usage dashboards which identify individual time and power usage to facilitate the ability to cost-split between groups. However, it should be noted that a continuity of visibility risks can give rise to surveillance anxiety—a tension well-documented in CSCW research \cite{Erickson2000SocialTranslucence}. While translucence can promote accountability, it can also heighten awareness of being watched. Effective design may therefore need to offer adjustable transparency levels, ensuring that coordination support does not become oppressive oversight.

In practice, default translucence levels should be tied to both role and sharing type. For owner-paid casual accounts, a conservative default—showcasing activity presence (e.g., “someone is active”) but hiding detailed query histories—respects the least comfortable member, while allowing the owner to selectively mark specific threads as shared. In cost-splitting groups, the highest-value opportunities appear to be group-level dashboards that show aggregate usage and presence by default, with fine-grained history visibility controlled by the author of each thread. When members disagree about the gradation of translucence—for instance, some wanting detailed token statistics and others finding them intrusive—the interface can make these preferences visible (e.g., per-user visibility settings attached to dashboards) and prompt groups to renegotiate a baseline rather than silently imposing the most intrusive option.

These design choices generate benefits for users through providing concrete means to articulate and coordinate norms, thus playing a more active role within the functionality and production of the LLM. This, in turn, reduces the probability of conflict and increases the level of trust. This dynamic exemplifies the CSCW principle of social translucence \cite{Erickson2000SocialTranslucence}: when meaningful visibility is provided, communities regulate and reinforce their own practices. Thus, design becomes less about reconstructing norms than creating environments in which they can develop naturally, which could in turn produce more sustainable user interaction, particularly given the novelty of LLMs for many of its users. 

\subsection{Adding Visibility Gradation to Manage Linguistic Traces}
The uniqueness of LLM sharing lies in the persistence of cognitive traces, which position users simultaneously as both observers and the observed. Our study found that even the mere possibility of being seen altered how participants interacted with LLMs through tone and word choice—an observer effect resonant with Sartre’s notion of the gaze \cite{Sartre2003BeingNothingness}. For Sartre, the gaze signifies the experience of being objectified in another’s consciousness, triggered not only by direct eye contact but also indirect cues. In shared accounts, these cues manifested through the form of conversation logs, folder structures, and token usage. Through these traces, users both interpreted others’ actions and anticipated judgments of their own, shifting between self-censorship out of discomfort and strategic disclosure seeking recognition.

This tension echoes extensive HCI privacy research showing that privacy is often managed through situational boundary negotiation and audience management rather than binary secrecy. Prior work on self-censorship and imagined audiences in online spaces demonstrates how the mere possibility of being seen can reshape expression and lead to last-minute suppression or polishing of content \cite{Das2013SelfCensoring, Marwick2011ImaginedAudience, Sleeper2013FacebookCensorship}. Research on shared and intimate technologies likewise shows that people rely on tacit norms, selective deletion, and workarounds to maintain privacy when accounts or devices are co-used \cite{Lampinen2011DisclosureManagement,Matthews2017StoriesFromSurvivors, Sambasivan2018Privacy}. Our findings extend these accounts to shared LLM histories, where persistent conversational traces can expose not only outcomes but also uncertainty, affect, and reasoning-in-progress—making boundary work both more valuable and more vulnerable.

Simple privacy toggles may be insufficient to address these dynamics, where users might be inclined to partially share some of their information for selective sharing that supports coordination, inspiration, or relationship maintenance—without requiring full exposure. In this case, we propose that prompt visibility should be conceived as a spectrum rather than a binary. That is, from seen/unseen chats to a granular imposition of boundaries where users can already adjust disclosure through selective sharing or careful phrasing, current interfaces treat all prompts as equally exposed. Here, users define the scope and mode of disclosure for each prompt. Here, such controls enable consent across diverse relational settings—families, colleagues, or semi-anonymous groups—while reducing cycles of snooping and self-censorship.

To operationalize this principle, interfaces could offer a three-tier visibility system: private, group, public. In this case, by default, conversations are shown simply as summaries where full transcripts can be accessed on request. This "temporary sandbox mode" could allow chats that do not maintain persistent tracing, while selective anonymization preserves the opportunity for mutual learning without attribution anxieties. These features reframe visibility within a negotiable spectrum, shifting the observer effect toward collaboration and knowledge exchange.

Granular visibility controls may increase interface complexity through adding decision overhead to routine interactions. Designers may therefore want to consider defaults settings carefully, minimizing decision overhead and supporting boundary management. In this case, the observer effect cannot be eliminated, but is productively redirected. The anticipated benefits are threefold: first, public sharing of prompts can turn voyeuristic snooping into legitimate peer learning; second, lockable threads provide the ability to create private chats, which in turn reduces the need to engage in implicit LLM etiquette, which is often fragile; and third, visible and negotiable boundaries foster trust and the ability to regulate emotional exposure in shared LLM environments. As such, the emergence of norms is encouraged naturally to the benefit of all. 

To make visibility gradation workable in practice, agency also needs to be clearly divided. We suggest that thread authors control the visibility of their own conversations, while account owners or administrators control group-level defaults. Author-level “locks” (e.g., marking a thread as Private or anonymized) should override more open group defaults, so that individuals can protect sensitive reasoning even in highly translucent groups. When someone attempts to broaden access to a locked or anonymized thread—for example, moving it from Private to Group—the system should notify the original author and defer the change until they consent. As a rule of thumb for practitioners, groups can decide how visible the shared space should be overall, but individuals should retain veto power over the exposure of their own linguistic traces.

%% file: sections/06-limitations.tex
\section{LIMITATIONS AND FUTURE WORK}
This study has several limitations. First, our survey and interview data relied on self-reports, which may differ from actual behavior, especially on sensitive topics where participants may underreport. To address this, we employed concrete scenarios and contextual probes to elicit more detailed responses, although discrepancies may still exist. Second, although our study addresses LLM services broadly, most cases concerned ChatGPT account sharing. We sought service diversity in recruitment; however, given ChatGPT's high adoption among domestic users, responses naturally leaned toward experiences of sharing ChatGPT accounts. Third, our cross-sectional design cannot capture temporal changes in norms and practices. We attempted to address this by differentiating between "past" and "current" experiences in interviews, but longitudinal insights remain beyond our scope.

Future work should involve cross-cultural comparisons and longitudinal studies to examine how implicit norms and observer effects evolve over time and across various contexts. To this end, it could employ more systematic sampling methods—such as stratified sampling or maximum variation sampling—to enhance the generalizability and external validity of the findings. Incorporating behavioral log data could complement self-reports and address their inherent limitations. Finally, the design ideas suggested here—such as supporting multiple profiles, separating conversation histories, and enabling selective visibility—should be tested and evaluated through user studies. These efforts will help LLM services move beyond single-user assumptions and foster sustainable support for multi-user environments.

%% file: sections/07-conclusion.tex
\section{CONCLUSION}
This study highlights that account sharing in LLM services is more than just about breaking rules or convenience; it is a social practice in which technologies built for individuals are used collectively. Account sharing is negotiated around access, responsibility, and visibility, leading to new norms and tensions. Our analysis identified four types of sharing based on whether the owner actively uses the account and whether subscription costs are divided. In the absence of clear rules, norms emerged and were interpreted differently by users, creating gaps that appear as observer effects. Users left traces of themselves and also held back their behavior because they were aware of being watched, experiencing both exposure and restraint. These findings have important implications for designing LLM services. By reframing account sharing around cognitive paths, voyeuristic learning, and intellectual vulnerability, our study extends prior research and identifies design opportunities unique to LLM-based sharing. Based on user experiences, our work emphasizes that LLM services may benefit from designs that support sustainability not only for individuals but also for collective use.

%% file: sections/Appendix-A.tex
\section{Appendix: Interview Protocol}

\setlength{\intextsep}{6pt plus 2pt minus 2pt}
\setlength{\textfloatsep}{6pt plus 2pt minus 2pt}
\setlength{\floatsep}{6pt plus 2pt minus 2pt}
\setlength{\abovecaptionskip}{4pt}
\setlength{\belowcaptionskip}{0pt}

\subsection{Introduction Script}
Thank you for taking the time to talk with me today. In this interview, I’m interested in your experiences using a shared LLM account. I’d like to hear how you share the account, manage your chat history, and what it’s like to see other people’s prompts and answers (or to imagine they might see yours). There are no right or wrong answers. I value your honest experiences and opinions. You can skip any question you prefer not to answer, and you can stop the interview at any time. With your permission, I’d like to record this conversation so I can accurately capture what you say. To begin, I’ll ask a few questions about how you share the account, then we’ll move into more specific examples and feelings.

\subsection{Context of Account Sharing and Relationship Structure}
First, I’d like to understand the basic context of the shared account you use.

\begin{table}[H]
\centering
\scriptsize
\setlength{\tabcolsep}{2pt}
 \caption{Context of Account Sharing and Relationship Structure}
\begingroup
 \renewcommand{\arraystretch}{0.8}
  \label{tab:Table 3}
  \input{tables/interview-protocol2} 
\endgroup
\end{table}

\subsection{Prompt History Management and Strategies}
Next, I’d like to ask about how you manage your conversation history or prompts within this shared account.

\begin{table}[H]
\centering
\scriptsize
\setlength{\tabcolsep}{2pt}
 \caption{Prompt History Management and Strategies}
\begingroup
\renewcommand{\arraystretch}{0.8}
  \label{tab:Table 4}

\input{tables/interview-protocol3} 
\endgroup
\end{table}

\subsection{Implicit Norms and Discomforts}
Now I’d like to talk about any unspoken rules or uncomfortable moments that have come up while sharing the account.

\begin{table}[H]
\centering
\scriptsize
\setlength{\tabcolsep}{2pt}
 \caption{Implicit Norms and Discomforts}
 \renewcommand{\arraystretch}{0.8}
  \label{tab:Table 5}
  \input{tables/interview-protocol4} 
\end{table}

\subsection{Encountering Others’ Records: Experiences and Emotions}
Next, I want to focus on what it’s like to encounter other people’s prompts and conversations in the shared account.
\begin{table}[H]
\centering
\scriptsize
\setlength{\tabcolsep}{2pt}
 \caption{Encountering Others’ Records: Experiences and Emotions}
\renewcommand{\arraystretch}{0.8}
  \label{tab:Table 6}
  \input{tables/interview-protocol5} 
\end{table}

\subsection{Indirect Communication Through the LLM as a Tool}

\textbf{Perceiving and Interpreting Others}

Now I’d like to talk about whether the shared account ever feels like a way to get to know the other users indirectly.

\begin{table}[H]
\centering
\scriptsize
\setlength{\tabcolsep}{2pt}
 \caption{Perceiving and Interpreting Others}
\renewcommand{\arraystretch}{0.8}
  \label{tab:Table 7}
  \input{tables/interview-protocol6} 
\end{table}

\textbf{Changes in Self‑Presentation and Awareness}

I’d also like to ask how your awareness of being seen might change how you present yourself in prompts.

\begin{table}[H]
\centering
\scriptsize
\setlength{\tabcolsep}{2pt}
 \caption{Changes in Self‑Presentation and Awareness}
\renewcommand{\arraystretch}{0.8}
  \label{tab:Table 8}
  \input{tables/interview-protocol7} 
\end{table}

\textbf{Indirect Interaction via the LLM}

Finally, I want to ask whether the LLM ever feels like a medium for indirect interaction with other account sharers.

\begin{table}[H]
\centering
\scriptsize
\setlength{\tabcolsep}{2pt}
 \caption{Indirect Interaction via the LLM}
\renewcommand{\arraystretch}{0.8}
  \label{tab:Table 9}
  \input{tables/interview-protocol8} 
\end{table}

\subsection{Closing Questions}
\begin{itemize}
    \item Is there anything about sharing an LLM account that we haven’t talked about but that feels important to you?
    \item If you could change anything about how shared accounts or histories work, what would you change?
    \item Do you have any questions for me about the study?
\end{itemize}
Thank you again for sharing your experiences. Your insights are very valuable for understanding how people use LLMs together.

%% file: tables/interview-protocol2.tex
\begin{tabular}{p{0.26\linewidth} p{0.70\linewidth}}
\toprule
\textbf{Category} & \textbf{Question} \\
\midrule
Account sharing timeline & Since when have you been sharing this LLM account? \\
Relationship with other users & With whom do you share this account? How would you describe your relationship with them? \\
Account ownership & Did you create this account yourself, or were you invited to join someone else's account? \\
Trigger for sharing & Was there any particular occasion or reason that led you to start sharing this account? \\
Cost and decision-making & How do you handle any costs associated with this account (if there are any)? \\
Roles in the sharing arrangement & Are there any implicitly agreed roles among the people who share this account? What are they? \\
\bottomrule
\end{tabular}

%% file: tables/interview-protocol3.tex
\begin{tabular}{p{0.26\linewidth} p{0.70\linewidth}}
\toprule
\textbf{Category} & \textbf{Question} \\
\midrule
Personal management of records & Within the shared account, do you manage your own conversation history separately in any way?  \\
Criteria for keeping or deleting history & Do you have any criteria for when to keep a chat record versus when to delete it? \\
Awareness of being seen by others & When you leave a record in the shared account, are you aware that other people might see it? How does that awareness affect what you do? \\
Etiquette or manners in history management & Do you have any personal standards of “etiquette” or “good manners” when it comes to managing the shared history? \\
Marking or finding one’s own records & Do you use any methods to distinguish your own records or to find them later easily? \\
Awareness of who used the account and when & Does the account have any structure that lets you know who used it and when? \\
\bottomrule
\end{tabular}

%% file: tables/interview-protocol4.tex
\begin{tabular}{p{0.26\linewidth} p{0.70\linewidth}}
\toprule
\textbf{Category} & \textbf{Question} \\
\midrule
Unspoken rules & Are there any “unspoken rules” that everyone seems to follow when using the shared account, even though you never explicitly talked about them?  \\
Areas of caution & Have you felt any need to be careful about certain things—such as browsing the history, the topics you ask about, or the time you use the account? \\
Perceptions of polite behavior & What kinds of behaviors do you consider to be “polite” or appropriate for someone who shares this account? \\
Moments of discomfort & Have there been any specific moments when sharing the account felt uncomfortable for you? \\
Coping and resolution &  In that situation, how did you respond or try to resolve the discomfort? \\
\bottomrule
\end{tabular}

%% file: tables/interview-protocol5.tex
\begin{tabular}{p{0.26\linewidth} p{0.70\linewidth}}
\toprule
\textbf{Category} & \textbf{Question} \\
\midrule
Exposure to others’ prompts & Have you ever read other people’s chat records in the shared account? What kinds of content did you see?  \\
Emotional reactions & How did you feel when you saw those records? \\
Subsequent actions & After seeing someone else’s record, did you take any actions? \\
Influence on one’s own prompting & Has seeing other people’s prompts or conversations ever influenced how you use prompts yourself? \\
Sense that others are viewing your records &  On the other side, have you ever felt that someone else was looking at your records? \\
Imagined reactions to others viewing your records &  When you imagine someone looking at your conversations, what kinds of thoughts or feelings come up? \\
Behavioral changes after noticing visibility &  After becoming aware that others might be viewing your records, have you ever changed or deleted content in response? \\
\bottomrule
\end{tabular}

%% file: tables/interview-protocol6.tex
\begin{tabular}{p{0.26\linewidth} p{0.70\linewidth}}
\toprule
\textbf{Category} & \textbf{Question} \\
\midrule
Glimpses into others’ minds & When using the shared account, have you ever felt that you were getting a glimpse into another person’s way of thinking or their interests?  \\
Realization of “how this person thinks” & Was there a particular prompt or response where you thought, “Ah, so this is how this person thinks”? \\
Impact on sense of “sharing” the account & How did those experiences affect your sense of “sharing” the account with others? \\
\bottomrule
\end{tabular}

%% file: tables/interview-protocol7.tex
\begin{tabular}{p{0.26\linewidth} p{0.70\linewidth}}
\toprule
\textbf{Category} & \textbf{Question} \\
\midrule
Consciousness of being read by others & Have you ever thought, “Someone might read this later, so I should write it better”?  \\
Situations where this awareness was strong & Can you describe a specific situation or example where you became especially conscious that your record might be seen by others? \\

\bottomrule
\end{tabular}

%% file: tables/interview-protocol8.tex
\begin{tabular}{p{0.26\linewidth} p{0.70\linewidth}}
\toprule
\textbf{Category} & \textbf{Question} \\
\midrule
Moments of indirect communication & Have you ever felt that you were indirectly communicating with someone else through the LLM? \\
Continuity of “conversation” via prompts & Even though it’s a shared account, have there been times when it felt like a conversation was continuing across different people’s questions? \\
\bottomrule
\end{tabular}

%% file: sections/Appendix-B.tex
\section{Appendix: Interview Codebook}

\subsection{Theme 1. Types of LLM Account Sharing}
\vspace{-0.4\baselineskip}
\begingroup
\setlength{\intextsep}{4pt}
\setlength{\textfloatsep}{4pt}
\setlength{\floatsep}{4pt}
\setlength{\abovecaptionskip}{2pt}
\setlength{\belowcaptionskip}{0pt}
\begin{table}[H]
\centering
\scriptsize
\setlength{\tabcolsep}{2pt}
 \caption{Types of LLM Account Sharing}
\renewcommand{\arraystretch}{0.8}
  \label{tab:Table 10}
  \input{tables/interview-codebook1} 
\end{table}
\endgroup
\subsection{Theme 2. Coordination Norms: Content, Types, and Fragility}
\vspace{-0.4\baselineskip}
\begingroup
\setlength{\intextsep}{4pt}
\setlength{\textfloatsep}{4pt}
\setlength{\floatsep}{4pt}
\setlength{\abovecaptionskip}{2pt}
\setlength{\belowcaptionskip}{0pt}
\begin{table}[H]
\centering
\scriptsize
\setlength{\tabcolsep}{2pt}
 \caption{Norm Content}
\renewcommand{\arraystretch}{0.8}
  \label{tab:Table 11}
  \input{tables/interview-codebook2a} 
\end{table}
\endgroup

\begin{table}[H]
\centering
\scriptsize
\setlength{\tabcolsep}{2pt}
 \caption{Norm Types}
\renewcommand{\arraystretch}{0.8}
  \label{tab:Table 12}
  \input{tables/interview-codebook2b} 
\end{table} 

\begin{table}[H]
\centering
\scriptsize
\setlength{\tabcolsep}{2pt}
 \caption{Norm Fragility and Violations}
\renewcommand{\arraystretch}{0.8}
  \label{tab:Table 13}
  \input{tables/interview-codebook2c} 
\end{table}

\subsection{Theme 3. Observer‑as‑Actor: Interpreting Others’ Traces and Acting on Them}

\begin{table}[H]
\centering
\scriptsize
\setlength{\tabcolsep}{2pt}
 \caption{Intentional Snooping}
\renewcommand{\arraystretch}{0.8}
  \label{tab:Table 14}

\input{tables/interview-codebook3a} 
\end{table}

\begin{table}[H]
\centering
\scriptsize
\setlength{\tabcolsep}{2pt}
 \caption{Unintentional Snooping}
\renewcommand{\arraystretch}{0.8}
  \label{tab:Table 15}
  \input{tables/interview-codebook3b} 
\end{table}

\begin{table}[H]
\centering
\scriptsize
\setlength{\tabcolsep}{2pt}
 \caption{Subsequences of Snooping}
\renewcommand{\arraystretch}{0.8}
  \label{tab:Table 16}
  \input{tables/interview-codebook3c} 
\end{table}

\subsection{Theme 4. Observer‑as‑Target: Being Seen and Managing Self‑Presentation}

\begin{table}[H]
\centering
\scriptsize
\setlength{\tabcolsep}{2pt}
 \caption{Awareness of potential observation}
\renewcommand{\arraystretch}{0.8}
  \label{tab:Table 17}
  \input{tables/interview-codebook4a} 
\end{table}

\begin{table}[H]
\centering
\scriptsize
\setlength{\tabcolsep}{2pt}
 \caption{Consequences of the awareness}
\renewcommand{\arraystretch}{0.8}
  \label{tab:Table 18}
  \input{tables/interview-codebook4b} 
\end{table}

\subsection{Theme 5. Recursive Cycle and Outcomes of Mutual Observation}
\begin{table}[H]
\centering
\scriptsize
\setlength{\tabcolsep}{2pt}
 \caption{Recursive Cycle and Outcomes of Mutual Observation}
\renewcommand{\arraystretch}{0.8}
  \label{tab:Table 19}
  \input{tables/interview-codebook5} 
\end{table}

%% file: tables/interview-codebook1.tex
\begin{tabular}{p{0.23\linewidth} p{0.73\linewidth}}
\toprule
\textbf{Code} & \textbf{Description(s)} \\
\midrule
Owned–Casual Sharing & Account owner actively uses the LLM and lets known others (often family or friends) use the same account without cost sharing. Owner typically manages payment, security, and settings, and sharing is framed as a favor or relational responsibility. Norms are often conveyed informally (e.g., over meals or chats), with the owner sometimes setting boundaries unilaterally. \\
Ownerless–Casual Sharing & A company or organization pays for an account that individual employees use, while the formal “owner” (the organization) does not personally use it. Management tends to be loose; formal rules (e.g., policy documents) exist but often have weak enforcement, and practical norms are negotiated in everyday work channels (Slack, internal messengers, etc.). \\
Owned–Split Sharing & Account owner both uses the LLM and splits subscription costs with peers (friends, colleagues, acquaintances). There are explicit arrangements around payment and periodic reminders, but many usage rules remain minimal or implicit to avoid social friction. Members often coordinate via separate group chats dedicated to money transfers and credentials. \\
Ownerless–Split Sharing & Participants share a subscription through an intermediary platform (e.g., group‑buy/“sharing” services) that manages accounts and billing. Co‑users are usually strangers or semi‑anonymous. There are no direct communication channels; norms (e.g., folder conventions, “don’t look at others’ chats”) are inferred from existing traces and quietly adopted, not negotiated. \\
\bottomrule
\end{tabular}

%% file: tables/interview-codebook2a.tex
\begin{tabular}{p{0.23\linewidth} p{0.73\linewidth}}
\toprule
\textbf{Code} & \textbf{Description(s)} \\
\midrule
Privacy Norm: Non‑Peeking into Others’ Chats & Any reference to an (often idealized) rule that one “should not” read others’ chat histories, folders, or projects. Includes statements like “don’t pry,” “don’t look at others’ conversations,” or “respect their privacy,” regardless of whether people actually follow it. \\
Privacy Norm: Self‑Management of Sensitive Content & Segments where participants describe expectations that each user should delete or avoid entering their own sensitive information (e.g., “if you ask a sensitive question, delete it yourself”). Emphasizes personal responsibility for minimizing exposure to co‑users. \\
Boundary Norm: Not Deleting or Editing Others’ Work & Mentions of avoiding deletion or alteration of others’ sessions, projects, or folders without permission. Includes statements like “don’t erase the session history without checking” or “don’t touch another person’s project.” \\
Boundary Norm: Keeping Sessions/Folders Separate & References to separating traces by person—creating separate sessions, folders, or project structures to avoid mixing work. Includes expectations like “ask questions only in your own folder” or “create new chats in your own space.” \\
Access Norm: Avoiding Simultaneous Use & Mentions of informal rules to avoid logging in at the same time, or coordinating to prevent slowdowns/lockouts. Includes talk about taking turns, respecting others’ “turn,” or checking if someone is using the account before starting a session. \\
Access Norm: Reasonable or Non‑Wasteful Use & Segments where participants talk about “not wasting tokens,” “using it reasonably,” or “if you don’t like the rule, get your own subscription.” Captures resource‑conservation attitudes and moralizing of overuse. \\
\bottomrule
\end{tabular}

%% file: tables/interview-codebook2b.tex
\begin{tabular}{p{0.23\linewidth} p{0.73\linewidth}}
\toprule
\textbf{Code} & \textbf{Description(s)} \\
\midrule
Explicit Co‑Negotiated Norms & Rules that are explicitly discussed and agreed among members as peers (e.g., in a group chat or meeting). Used when participants describe having talked about “how we’ll use it,” “who can delete what,” or “how to split capacity” as a joint decision for fairness. \\
Explicit Owner‑Imposed Norms & Rules declared unilaterally by the owner, often justified by payment or account control (e.g., “it’s my account so I keep the right to delete anything”). Includes segments where owners frame rule‑setting as part of their responsibility and privilege as payer. \\
Implicit Norms Learned by Observing Collective Practice & Norms inferred by watching how other people seem to use the account—e.g., copying existing folder structures, matching others’ style of naming, or “just doing what everyone seems to be doing.” No explicit conversation is reported; learning is through observation and mimicry. \\
Implicit Norms Individually Defined and Projected onto Others & Segments where a participant describes their own personal rule (“I never put anything personal here,” “I always delete everything”) and assumes or hopes others do the same. These norms are not actually negotiated but are treated as if they were shared common sense. \\
\bottomrule
\end{tabular}

%% file: tables/interview-codebook2c.tex
\begin{tabular}{p{0.23\linewidth} p{0.73\linewidth}}
\toprule
\textbf{Code} & \textbf{Description(s)} \\
\midrule
Norm Ambiguity and Misalignment & Evidence that people in the same account interpret “the rules” differently (e.g., one spouse says “you can look,” another says “you shouldn’t”). Includes vague concepts like “sensitive information” that participants acknowledge can mean different things to different people, and agreements or responsibilities that are often invisible or easily forgotten. \\
Privacy Norm Violation: Snooping / Peeking & Any admission or description of looking at others’ chat histories, folders, or usage despite nominal norms against it. Includes both deliberate snooping (“I check my sibling’s usage”) and reluctant confession (“I know I shouldn’t, but I looked”). \\
Enforcement Challenges and Privacy Paradox & Mentions of the structural difficulty of enforcing privacy norms: that checking compliance often requires seeing others’ content, which itself can violate privacy. Includes awareness that snooping leaves few traces, making violations low‑risk and hard to sanction. \\
\bottomrule
\end{tabular}

%% file: tables/interview-codebook3a.tex
\begin{tabular}{p{0.23\linewidth} p{0.73\linewidth}}
\toprule
\textbf{Code} & \textbf{Description(s)} \\
\midrule
Curiosity‑Driven Snooping & Participants describe opening others’ chats or folders “just to see” what they are doing, without a specific practical purpose. Includes language of voyeurism, fascination, or entertainment (“like peeking into how others live”). \\
Welfare or Care‑Driven Monitoring & Segments where people look at others’ traces to check on their state, wellbeing, or competence (e.g., parents checking a child’s usage, siblings monitoring each other). Snooping is framed as caretaking or responsibility. \\
Performance/Access Monitoring & Looking at history or usage to see how heavily others are using the account, whether they might be exceeding limits, or whether the system will be slow. Includes checking project folders or usage to manage congestion or fairness. \\
Learning‑Oriented Observation & Observing others’ prompts or conversations specifically to learn phrasing, strategies, or domain knowledge. Includes copying prompting styles (“I started writing ‘I’m a UX major…’ after seeing him do it”) or reusing Q\&A as templates for one’s own work. \\
\bottomrule
\end{tabular}

%% file: tables/interview-codebook3b.tex
\begin{tabular}{p{0.23\linewidth} p{0.73\linewidth}}
\toprule
\textbf{Code} & \textbf{Description(s)} \\
\midrule
Unintentional Snooping & Cases where someone lands on others’ chats by accidental clicks or page transitions (“I clicked and saw the last one or two chats, that’s it”). They didn’t seek it out but still briefly view traces. \\
\bottomrule
\end{tabular}

%% file: tables/interview-codebook3c.tex
\begin{tabular}{p{0.23\linewidth} p{0.73\linewidth}}
\toprule
\textbf{Code} & \textbf{Description(s)} \\
\midrule
Emotional Reactions as Observer & Emotions felt after seeing others’ records, such as guilt, discomfort, amusement, empathy, or resentment. Includes strong reactions to private content (e.g., a child writing about feelings after being scolded) and mixed feelings about having seen it. \\
Acting on Observed Traces (Online/Offline) & Behavior changes after viewing others’ records: joking about them, confronting the person, using the content in work, or referencing them in conversation. Includes using others’ Q\&A as material (e.g., translating or reworking answers) and social interactions triggered by shared traces. \\
\bottomrule
\end{tabular}

%% file: tables/interview-codebook4a.tex
\begin{tabular}{p{0.23\linewidth} p{0.73\linewidth}}
\toprule
\textbf{Code} & \textbf{Description(s)} \\
\midrule
Perceived Visibility and Security Concerns & Participants explicitly express worry that co‑users can see what they ask (“what if someone sees what I’ve been asking?”). Captures the moment when they shift from feeling like a private user to feeling observed and potentially judged. \\
\bottomrule
\end{tabular}

%% file: tables/interview-codebook4b.tex
\begin{tabular}{p{0.23\linewidth} p{0.73\linewidth}}
\toprule
\textbf{Code} & \textbf{Description(s)} \\
\midrule
Pre‑Emptive Censorship (Before Writing) & Self‑censorship that happens before sending a prompt: avoiding certain topics, toning down intimacy, or using more formal/polite language because they anticipate others might read it. Includes reflections like “I feel less free” or “I wouldn’t ask MBTI‑style questions here.” \\
Post‑Hoc Concealment (After Writing) & Managing traces after the fact: deleting chats, editing content, using “temporary” modes, or creating separate free accounts for personal/sensitive questions. Applied when participants describe actively cleaning up or relocating certain topics. \\
Strategic Positive Self‑Presentation & Deliberately leaving visible traces to look competent, diligent, or thoughtful to imagined co‑users. Examples include polishing English prompts, showcasing specific questions, or explicitly saying they want others to see their good usage so it “rubs off in a good way.” \\
Relationship‑Specific Calibration of Disclosure & Different self‑presentation strategy depending on who is sharing the account (e.g., more self‑conscious with close friends or siblings, more relaxed with strangers in group‑buy settings). Used when participants compare comfort levels across relational contexts. \\
Instrumental Sharing for Collective Benefit & Segments where participants leave traces intentionally because they believe others can reuse them (e.g., prompts as shared resources, reusable Q\&A). Here, visibility is framed as a benefit to the group rather than a pure privacy risk. \\
\bottomrule
\end{tabular}

%% file: tables/interview-codebook5.tex
\begin{tabular}{p{0.23\linewidth} p{0.73\linewidth}}
\toprule
\textbf{Code} & \textbf{Description(s)} \\
\midrule
Mutual Surveillance & Explicit recognition that if you can see others, they can see you, producing a mirror‑like feeling (“every time I look at someone else’s record, it feels like a mirror”). Captures the idea of users cycling between observer and observed roles. \\
Shared Account as Performance Stage & Descriptions of the shared account as a place where users “perform” a certain persona—carefully editing wording, deleting traces to avoid particular impressions, or curating what remains. Includes talk about not wanting others to think “she only works” or “she doesn’t know this,” leading to extensive clean‑up. \\
Shared Account as Learning and Inspiration Space & Segments emphasizing the upside of visibility: using others’ traces for learning prompting strategies, discovering new use cases, or reducing duplicated effort (“I can get information without asking the question myself”). \\
Shared Account as Relationship and Emotion Resource & Moments where shared traces enable understanding others’ situations, feelings, or interests (e.g., learning a sibling has gastroenteritis, understanding partner’s side of a conflict). Includes indirect emotional communication via the LLM history. \\
Spillover into Direct Interpersonal Interaction & Cases where content originally written “for the LLM” becomes part of human–human conversations (e.g., partner confronting the phrasing of a prompt about a fight). Used when LLM logs are explicitly referenced in later chats or conflicts. Includes situations where participants write to the LLM with an anticipated human observer. \\
Exit or Parallel Use Due to Burden & Participants describe avoiding the shared account for personal things, opening a separate free account, or considering leaving entirely because the psychological burden of being watched is too high. \\
\bottomrule
\end{tabular}

%% file: references/bibliography.bib
@misc{save-streaming,
 author = {Rick Boida, Kourtnee Jackson},
 title = {{10 Things You Can Do To Save Money on Streaming}},
 howpublished = "\url{https://www.cnet.com/tech/services-and-software/10-things-you-can-do-to-save-money-on-streaming/}", 
 year = {2023},
 note = "[Online; accessed August-30-2025]"
}

@misc{openaisharingpolicy,
  author = {OpenAI Account Sharing Policy},
  title = {{How to Share a ChatGPT Plus Account Safely with Friends, Teams, or Group Buys}},
  howpublished = "\url{https://help.openai.com/en/articles/10471989-openai-account-sharing-policy/}",
  year = {2025}, 
  note = "[Online; accessed August-30-2025]"
}

@misc{openaibusiness,
  author = {What is ChatGPT Business?},
  title = {{How to Share a ChatGPT Plus Account Safely with Friends, Teams, or Group Buys}},
  howpublished = "\url{https://help.openai.com/en/articles/8792828-what-is-chatgpt-business/}",
  year = {2025}, 
  note = "[Online; accessed August-30-2025]"
}

@article{Dourish2003,
author = {Dourish, Paul},
title = {The Appropriation of Interactive Technologies: Some Lessons from Placeless Documents},
year = {2003},
journal = {Computer Supported Cooperative Work (CSCW)},
url = {https://doi.org/10.1023/A:1026149119426},
doi = {10.1023/A:1026149119426}
}

@article{Schmidt1992,
author = {Schmidt, Kjeld and Bannon, Liam},
title = {Taking CSCW seriously},
year = {1992},
journal = {Computer Supported Cooperative Work (CSCW)},
url = {https://doi.org/10.1007/BF00752449},
doi = {10.1007/BF00752449}
}

@inproceedings{Boardman2004PIM,
  author    = {Richard Boardman and Martina Angela Sasse},
  title     = {{'Stuff goes into the computer and doesn’t come out': A cross-tool study of personal information management}},
  booktitle = {Proceedings of the SIGCHI Conference on Human Factors in Computing Systems (CHI '04)},
  year      = {2004},
  pages     = {583--590},
  publisher = {ACM},
  address   = {New York, NY, USA},
  doi       = {10.1145/985692.985765},
  isbn      = {1-58113-702-8}
}

@incollection{Stevens2009AppropriationInfrastructure,
  author    = {Gunnar Stevens and Volkmar Pipek and Volker Wulf},
  title     = {Appropriation Infrastructure: Supporting the Design of Usages},
  booktitle = {End-User Development. IS-EUD 2009. Lecture Notes in Computer Science},
  editor    = {Volkmar Pipek and Mary Beth Rosson and Boris de Ruyter and Volker Wulf},
  series    = {Lecture Notes in Computer Science},
  volume    = {5435},
  pages     = {50--69},
  year      = {2009},
  publisher = {Springer, Berlin, Heidelberg},
  doi       = {10.1007/978-3-642-00427-8_4},
  isbn      = {978-3-642-00425-4, 978-3-642-00427-8}
}

@inproceedings{Salovaara2006StudyingAppropriation,
  author    = {Antti Salovaara},
  title     = {Studying Appropriation of Mobile Technologies},
  booktitle = {MobileHCI 2006 Doctoral Consortium},
  year      = {2006},
  address   = {Espoo, Finland},
  note      = {MobileHCI Doctoral Consortium Paper}
}

@article{Salovaara2011EverydayAppropriations,
  author    = {Antti Salovaara and Sacha Helfenstein and Antti Oulasvirta},
  title     = {Everyday Appropriations of Information Technology: A Study of Creative Uses of Digital Cameras},
  journal   = {Journal of the American Society for Information Science and Technology},
  year      = {2011},
  volume    = {62},
  number    = {12},
  pages     = {2347--2363},
  doi       = {10.1002/asi.21643},
}

@article{Wakkary2008AspectsOfEverydayDesign,
  author    = {Ron Wakkary and Leah Maestri},
  title     = {Aspects of Everyday Design: Resourcefulness, Adaptation, and Emergence},
  journal   = {International Journal of Human–Computer Interaction},
  year      = {2008},
  volume    = {24},
  number    = {5},
  pages     = {478--491},
  doi       = {10.1080/10447310802142276},
}

@article{BanslerHavn2006Sensemaking,
  author    = {J{\o}rgen P. Bansler and Erling C. Havn},
  title     = {Sensemaking in Technology-Use Mediation: Adapting Groupware Technology in Organizations},
  journal   = {Computer Supported Cooperative Work},
  year      = {2006},
  volume    = {15},
  number    = {1},
  pages     = {55--91},
  doi       = {10.1007/s10606-005-9012-x},
  publisher = {Springer}
}

@book{Rogers2011HCITheory,
  author    = {Yvonne Rogers},
  title     = {HCI Theory: Classical, Modern, and Contemporary},
  year      = {2011},
  publisher = {Morgan \& Claypool},
  series    = {Synthesis Lectures on Human-Centered Informatics},
  volume    = {5},
  number    = {2},
  pages     = {1--129},
  doi       = {10.2200/S00318ED1V01Y201105HCI011},
  isbn      = {9781608457020}
}

@inproceedings{ObadaObieh2020Burden,
  author    = {Borke Obada-Obieh and Yue Huang and Konstantin Beznosov},
  title     = {The Burden of Ending Online Account Sharing},
  booktitle = {Proceedings of the 2020 CHI Conference on Human Factors in Computing Systems},
  year      = {2020},
  pages     = {1--13},
  publisher = {ACM}
}

@inproceedings{Hayashi2012Goldilocks,
  author    = {Eiji Hayashi and Oriana Riva and Karin Strauss and A. J. Bernheim Brush and Stuart E. Schechter},
  title     = {Goldilocks and the Two Mobile Devices: Going Beyond All--Or--Nothing Access to a Device’s Applications},
  booktitle = {Proceedings of the Eighth Symposium on Usable Privacy and Security (SOUPS)},
  year      = {2012},
  pages     = {2:1--2:11},
  publisher = {ACM},
  doi       = {10.1145/2335356.2335359},
  address   = {Washington, DC, USA}
}

@inproceedings{BrushInkpen2007YoursMineOurs,
  author    = {A. J. Bernheim Brush and Kori M. Inkpen},
  title     = {Yours, Mine and Ours? Sharing and Use of Technology in Domestic Environments},
  booktitle = {Proceedings of the 9th International Conference on Ubiquitous Computing (UbiComp 2007)},
  editor    = {John Krumm and Gregory D. Abowd and Aruna Seneviratne and Thomas Strang},
  series    = {Lecture Notes in Computer Science},
  volume    = {4717},
  pages     = {109--126},
  year      = {2007},
  publisher = {Springer, Berlin, Heidelberg},
  doi       = {10.1007/978-3-540-74853-3_7}
}

@incollection{FrohlichKraut2003SocialContext,
  author      = {David M. Frohlich and Robert E. Kraut},
  title       = {The Social Context of Home Computing},
  booktitle   = {Inside the Smart Home},
  editor      = {Rob Harper},
  publisher   = {Springer-Verlag},
  address     = {London},
  pages       = {127--162},
  year        = {2003},
  note        = {Chapter 8},
  doi         = {10.1007/1-85233-854-7_8}
}

@inproceedings{Matthews2016EverydayDeviceSharing,
  author    = {Tara Matthews and Kerwell Liao and Anna Turner and Marianne Berkovich and Rob Reeder and Sunny Consolvo},
  title     = {{She’ll Just Grab Any Device That’s Closer}: A Study of Everyday Device \& Account Sharing in Households},
  booktitle = {Proceedings of the 2016 ACM Conference on Human Factors in Computing Systems (CHI '16)},
  year      = {2016},
  pages     = {5921--5932},
  publisher = {ACM},
  doi       = {10.1145/2858036.2858051}
}

@inproceedings{Jacobs2016CaringSharing,
  author    = {Maureen Jacobs and Tanya D. Zuk and Konstantin Beznosov and Lorie Faith Cranor and Serge Egelman and Jens Grossklags and Julie S. Downs and Mary Ellen Zurko},
  title     = {Caring About Sharing: Couples’ Practices in Single User Device Access},
  booktitle = {Proceedings of the 19th ACM Conference on Computer-Supported Cooperative Work and Social Computing (CSCW '16)},
  year      = {2016},
  pages     = {37--50},
  publisher = {ACM},
  address   = {New York, NY, USA},
  doi       = {10.1145/2818048.2820075},
  isbn      = {978-1-4503-3592-8}
}

@article{Blythe2013Circumvention,
  author    = {John M. Blythe and Lynne Coventry and Judith Masthoff},
  title     = {Circumvention of Security: Good Users Do Bad Things},
  journal   = {IEEE Security \& Privacy},
  year      = {2013},
  volume    = {11},
  number    = {5},
  pages     = {80--83},
  publisher = {IEEE},
  doi       = {10.1109/MSP.2013.111},
  issn      = {1540-7993}
}

@inproceedings{Song2019NormalEasy,
  author    = {Yixin Song and Lujo Bauer and Nicolas Christin and Lorrie Faith Cranor and Apu Kapadia},
  title     = {"Normal and Easy": Account Sharing Practices in the Workplace},
  booktitle = {Fifteenth Symposium on Usable Privacy and Security (SOUPS 2019)},
  year      = {2019},
  pages     = {account-sharing-2019},  
  publisher = {USENIX Association},
  address   = {Santa Clara, CA, USA},
  url       = {https://www.usenix.org/conference/soups2019/presentation/song},
}

@article{AdamsSasse1999Users,
  author    = {Anne Adams and M. Angela Sasse},
  title     = {Users are not the enemy},
  journal   = {Communications of the ACM},
  year      = {1999},
  volume    = {42},
  number    = {12},
  pages     = {40--46},
  publisher = {ACM},
  doi       = {10.1145/322796.322806},
  issn      = {0001-0782}
}

@inproceedings{Song2021IMightBeUsingHis,
  author    = {Ji Eun Song and Jaeyoun You and Joongseek Lee},
  title     = {{“I Might be Using His… But It Is Also Mine!”: Ownership and Control in Accounts Designed for Sharing}},
  booktitle = {Proceedings of the 2021 ACM Conference on Human Factors in Computing Systems (CHI '21)},
  year      = {2021},
  pages     = {1--15},
  publisher = {ACM},
  doi       = {10.1145/3411764.3445301},
  address   = {Yokohama, Japan}
}

@article{Prottoy2025IfWeHadTheOption,
  author    = {Hasan Mahmud Prottoy and Yaxing Yao and Foad Hamidi},
  title     = {“If We Had the Option”: Infrastructuring for Access to Online Subscription-Based Services in Bangladesh},
  journal   = {Proceedings of the ACM on Human-Computer Interaction},
  year      = {2025},
  volume    = {9},
  number    = {2},
  pages     = {Article CSCW112:1--28},
  doi       = {10.1145/3711010},
  publisher = {ACM},
  address   = {New York, NY, USA},
  month     = apr
}

@article{Erickson2000SocialTranslucence,
  author    = {Thomas Erickson and Wendy A. Kellogg},
  title     = {Social Translucence: An Approach to Designing Systems that Support Social Processes},
  journal   = {ACM Transactions on Computer-Human Interaction},
  year      = {2000},
  volume    = {7},
  number    = {1},
  pages     = {59--83},
  doi       = {10.1145/344949.345004},
  publisher = {ACM},
  issn      = {1073-0516}
}

@inproceedings{Dourish1992Awareness,
  author    = {Paul Dourish and Victoria Bellotti},
  title     = {Awareness and Coordination in Shared Workspaces},
  booktitle = {Proceedings of the 1992 ACM Conference on Computer-Supported Cooperative Work (CSCW '92)},
  year      = {1992},
  pages     = {107--114},
  publisher = {ACM},
  address   = {New York, NY, USA},
  doi       = {10.1145/143457.143468},
  isbn      = {0-89791-542-9}
}

@article{Koppel2015Workarounds,
  author    = {Ross Koppel and Sean M. Smith and Jim Blythe and Virginia K. Kothari},
  title     = {Workarounds to Computer Access in Healthcare Organizations: You Want My Password or a Dead Patient?},
  journal   = {Journal of the American Medical Informatics Association},
  year      = {2015},
  volume    = {22},
  number    = {6},
  pages     = {1165--1173},
  doi       = {10.1093/jamia/ocv070},
  publisher = {Oxford University Press},
  issn      = {1527-974X}
}

@inproceedings{Baclawski2018ObserverEffect,
  author    = {Kenneth Baclawski},
  title     = {The Observer Effect},
  booktitle = {Proceedings of the 2018 IEEE Conference on Cognitive and Computational Aspects of Situation Management (CogSIMA)},
  year      = {2018},
  pages     = {83--89},
  publisher = {IEEE},
  address   = {Boston, MA, USA},
  doi       = {10.1109/COGSIMA.2018.8423983}
}

@article{Fox2008ClinicalEstimationFetalWeightHawthorne,
  author    = {Nathan S. Fox and Jennifer S. Brennan and Stephen T. Chasen},
  title     = {Clinical Estimation of Fetal Weight and the Hawthorne Effect},
  journal   = {European Journal of Obstetrics \& Gynecology and Reproductive Biology},
  year      = {2008},
  volume    = {141},
  number    = {2},
  pages     = {111--114},
  doi       = {10.1016/j.ejogrb.2008.07.023},
  publisher = {Elsevier}
}

@article{LeCompte1982ReliabilityValidity,
  author    = {Margaret D. LeCompte and Jean J. Goetz},
  title     = {Problems of Reliability and Validity in Ethnographic Research},
  journal   = {Review of Educational Research},
  year      = {1982},
  volume    = {52},
  number    = {1},
  pages     = {31--60},
  doi       = {10.3102/00346543052001031},
  publisher = {SAGE Publications},
  issn      = {0034-6543}
}

@article{Oswald2014HandlingHawthorne,
  author    = {David Oswald and Fred Sherratt and Simon Smith},
  title     = {Handling the Hawthorne Effect: The Challenges Surrounding a Participant Observer},
  journal   = {Review of Social Studies},
  year      = {2014},
  volume    = {1},
  number    = {1},
  pages     = {53--73},
}

@article{CoombsSmith2003HawthorneEffect,
  author    = {Steven J. Coombs and Ian D. Smith},
  title     = {The Hawthorne Effect: Is it a Help or a Hindrance in Social Science Research?},
  journal   = {Change: Transformations in Education},
  year      = {2003},
  volume    = {6},
  number    = {1},
  pages     = {97--111},
  note      = {University of Sydney},
}

@article{ChiesaHobbs2008Hawthorne,
  author    = {Mecca Chiesa and Sandy Hobbs},
  title     = {Making Sense of Social Research: How Useful is the Hawthorne Effect?},
  journal   = {European Journal of Social Psychology},
  year      = {2008},
  volume    = {38},
  number    = {1},
  pages     = {67--74},
  doi       = {10.1002/ejsp.401},
  publisher = {John Wiley \& Sons}
}

@inproceedings{Park2018ShareAndShareAlike,
  author    = {Cheul Young Park and Cori Faklaris and Siyan Zhao and Alex Sciuto and Laura Dabbish and Jason I. Hong},
  title     = {Share and Share Alike? An Exploration of Secure Behaviors in Romantic Relationships},
  booktitle = {Proceedings of the Fourteenth Symposium on Usable Privacy and Security (SOUPS 2018)},
  year      = {2018},
  pages     = {83--102},
  address   = {Baltimore, MD, USA},
  publisher = {USENIX Association},
  url       = {https://www.usenix.org/conference/soups2018/presentation/park}
}

@inproceedings{Naveed2022PrivacyBeyondWEIRD,
  author    = {Sheza Naveed and Hamza Naveed and Mobin Javed and Maryam Mustafa},
  title     = {{``Ask this from the person who has private stuﬀ'': Privacy Perceptions, Behaviours and Beliefs Beyond W.E.I.R.D}},
  booktitle = {Proceedings of the 2022 CHI Conference on Human Factors in Computing Systems (CHI '22)},
  year      = {2022},
  pages     = {1--17},
  publisher = {ACM},
  address   = {New Orleans, LA, USA},
  doi       = {10.1145/3491102.3501883},
  isbn      = {978-1-4503-9157-3/22/04}
}

@inproceedings{Matthews2017StoriesFromSurvivors,
  author    = {Tara Matthews and Kathleen O'Leary and Anna Turner and Manya Sleeper and Jill Palzkill Woelfer and Martin Shelton and Cori Manthorne and Elizabeth F. Churchill and Sunny Consolvo},
  title     = {{Stories from Survivors: Privacy \& Security Practices when Coping with Intimate Partner Abuse}},
  booktitle = {Proceedings of the 2017 CHI Conference on Human Factors in Computing Systems (CHI '17)},
  year      = {2017},
  pages     = {2189--2201},
  publisher = {ACM},
  address   = {New York, NY, USA},
  doi       = {10.1145/3025453.3025875},
  isbn      = {978-1-4503-4655-9}
}

@article{Blom2005ContextualCultural,
  author    = {Jan Blom and Jan Chipchase and Jaakko Lehikoinen},
  title     = {Contextual and Cultural Challenges for User Mobility Research},
  journal   = {Communications of the ACM},
  year      = {2005},
  volume    = {48},
  number    = {7},
  pages     = {37--41},
  doi       = {10.1145/1070838.1070857},
  publisher = {ACM},
  issn      = {0001-0782}
}

@article{Hofstede2011Dimensionalizing,
  author    = {Geert Hofstede},
  title     = {Dimensionalizing Cultures: The Hofstede Model in Context},
  journal   = {Online Readings in Psychology and Culture},
  year      = {2011},
  volume    = {2},
  number    = {1},
  pages     = {8},
  doi       = {10.9707/2307-0919.1014},
  publisher = {International Association for Cross-Cultural Psychology},
  issn      = {2307-0919}
}

@techreport{Graham2012InterRater,
  author    = {Matthew Graham and Anthony Milanowski and Jackson Miller},
  title     = {Measuring and Promoting Inter-Rater Agreement of Teacher and Principal Performance Ratings},
  institution = {Regional Educational Laboratory (REL) Midwest, Institute of Education Sciences, U.S. Department of Education},
  year      = {2012},
  number    = {REL 2012--No. 002},
  address   = {Washington, DC},
  url       = {https://ies.ed.gov/ncee/edlabs/regions/midwest/pdf/REL_2012002.pdf}
}

@inproceedings{Gaver2003Ambiguity,
  author    = {William W. Gaver and Jacob Beaver and Steve Benford},
  title     = {Ambiguity as a Resource for Design},
  booktitle = {Proceedings of the SIGCHI Conference on Human Factors in Computing Systems (CHI '03)},
  year      = {2003},
  pages     = {233--240},
  publisher = {ACM},
  address   = {New York, NY, USA},
  doi       = {10.1145/642611.642653},
  isbn      = {1-58113-630-7}
}

@article{Tchounikine2017DesigningForAppropriation,
  author    = {Pierre Tchounikine},
  title     = {Designing for Appropriation: A Theoretical Account},
  journal   = {Human--Computer Interaction},
  year      = {2017},
  volume    = {32},
  number    = {4},
  pages     = {155--195},
  doi       = {10.1080/07370024.2016.1203263},
  publisher = {Taylor \& Francis}
}

@book{Sartre2003BeingNothingness,
  author    = {Jean-Paul Sartre},
  title     = {Being and Nothingness: An Essay on Phenomenological Ontology},
  year      = {2003},
  publisher = {Routledge},
  address   = {London, UK},
  translator = {Hazel E. Barnes},
  isbn      = {9780415278485},
  note      = {Original work published 1943}
}

@misc{gamsgo,
author = {Gamsgo},
  title = {{Account sharing Platform}},
  howpublished = "\url{https://www.gamsgo.com/}",
  year = {2025}, 
  note = "[Online; accessed August-30-2025]"
}

@inproceedings{Sailaja2022AccountSharing,
  author    = {Neelima Sailaja and Abigail Fowler},
  title     = {An Exploration of Account Sharing Practices on Media Platforms},
  booktitle = {Proceedings of the ACM International Conference on Interactive Media Experiences (IMX '22)},
  year      = {2022},
  pages     = {274--279},
  publisher = {ACM},
  address   = {New York, NY, USA},
  doi       = {10.1145/3505284.3529975},
  isbn      = {978-1-4503-9249-6}
}

@inproceedings{Sambasivan2018Privacy,
  author    = {Nithya Sambasivan and Rahul A. Deshmukh and Aditya Dhruv and Edmond W. Felten and Janice Tsai and Sunny Consolvo},
  title     = {``Privacy is not for Me, It's for Those Rich Women'': Performative Privacy Practices on Mobile Phones by Women in South Asia},
  booktitle = {Proceedings of the Fourteenth Symposium on Usable Privacy and Security (SOUPS 2018)},
  year      = {2018},
  pages     = {127--142},
  publisher = {USENIX Association},
  address   = {Baltimore, MD, USA},
  url       = {https://www.usenix.org/conference/soups2018/presentation/sambasivan}
}

@inproceedings{Das2013SelfCensoring,
  title = {Self-Censorship on Facebook},
  author = {Das, Sauvik and Kramer, Adam D.I.},
  booktitle = {Proc. 7th Int. AAAI Conference on Weblogs and Social Media (ICWSM)},
  pages = {120--127},
  year = {2013}
}

@inproceedings{Saha2024ObserverEffect,
  title = {Observer Effect in Social Media Use},
  author = {Saha, Koustuv and Gupta, Pranshu and Mark, Gloria and K{\i}c{\i}man, Emre and De Choudhury, Munmun},
  booktitle = {Proc. CHI Conference on Human Factors in Computing Systems},
  pages = {Paper 138, 19 pages},
  year = {2024},
  doi = {10.1145/3576840}
}

@article{Marder2016ChillingFacebook,
  title = {The Extended 'Chilling' Effect of Facebook: The Cold Reality of Ubiquitous Social Networking},
  author = {Marder, Ben and Joinson, Adam and Shankar, Avi and Houghton, David},
  journal = {Computers in Human Behavior},
  volume = {60},
  pages = {582--592},
  year = {2016},
  doi = {10.1016/j.chb.2016.02.097}
}

@inproceedings{Candello2019AudienceEffectCHI,
  title = {The Effect of Audiences on the User Experience with Conversational Interfaces in Physical Spaces},
  author = {Candello, Heloisa and Pinhanez, Claudio and Pichiliani, Mauro and Cavalin, Paulo and Figueiredo, Flavio and Vasconcelos, Marisa},
  booktitle = {Proc. CHI Conference on Human Factors in Computing Systems},
  pages = {Paper 90, 13 pages},
  year = {2019},
  doi = {10.1145/3290605.3300320}
}

@article{Marwick2011ImaginedAudience,
  title = {I tweet honestly, I tweet passionately: Twitter users, context collapse, and the imagined audience},
  author = {Marwick, Alice E. and boyd, danah},
  journal = {New Media \& Society},
  volume = {13},
  number = {1},
  pages = {114--133},
  year = {2011},
  doi = {10.1177/1461444810365313}
}

@inproceedings{Sleeper2013FacebookCensorship,
  title = {The post that wasn't: exploring self-censorship on Facebook},
  author = {Sleeper, Manya and Balebako, Rebecca and Das, Sauvik and McConahy, Amber L. and Wiese, Jason and Cranor, Lorrie F.},
  booktitle = {Proc. ACM CSCW Conference},
  pages = {793--802},
  year = {2013},
  doi = {10.1145/2441776.2441865}
}

@inproceedings{Birnholtz2012Visibility,
  title = {Do You See That I See? Effects of Perceived Visibility on Awareness Checking Behavior},
  author = {Birnholtz, Jeremy and Bi, Nanyi and Fussell, Susan},
  booktitle = {Proc. ACM CHI Conference on Human Factors in Computing Systems},
  pages = {1765--1774},
  year = {2012},
  doi = {10.1145/2207676.2208312}
}

@article{Mitchell2011ImpressionManagementHCI,
  title = {Does social desirability bias favor humans? Explicit–implicit evaluations of synthesized speech support a new HCI model of impression management},
  author = {Mitchell, Wade J. and Ho, Chin-Chang and Patel, Himalaya and MacDorman, Karl F.},
  journal = {Computers in Human Behavior},
  volume = {27},
  number = {1},
  pages = {402--412},
  year = {2011},
  doi = {10.1016/j.chb.2010.09.018}
}

@article{Obrenovic2014HawthorneHCI,
  title = {The Hawthorne studies and their relevance to HCI research},
  author = {Obrenović, Željko},
  journal = {Interactions},
  volume = {21},
  number = {6},
  pages = {46--51},
  year = {2014},
  doi = {10.1145/2674966}
}

@article{Dym2020NormVulnerability,
  title = {Social Norm Vulnerability and its Consequences for Privacy and Safety in an Online Community},
  author = {Brianna Dym and Casey Fiesler},
  journal = {Proceedings of the ACM on Human-Computer Interaction (PACMHCI)},
  volume = {4},
  number = {CSCW2},
  pages = {155:1--155:24},
  year = {2020},
  doi = {10.1145/3415226}
}

@article{Chandrasekharan2018RedditNormViolations,
  title = {The Internet's Hidden Rules: An Empirical Study of Reddit Norm Violations at Micro, Meso, and Macro Scales},
  author = {Eshwar Chandrasekharan and Mattia Samory and Shagun Jhaver and Hunter Charvat and Amy S. Bruckman and Cliff Lampe and Jacob Eisenstein and Eric Gilbert},
  journal = {Proceedings of the ACM on Human-Computer Interaction (PACMHCI)},
  volume = {2},
  number = {CSCW},
  pages = {32:1--32:25},
  year = {2018},
  doi = {10.1145/3274301}
}

@inproceedings{Seering2017TwitchNorms,
  title = {Shaping Pro- and Anti-Social Behavior on Twitch Through Moderation and Example-Setting},
  author = {Joseph Seering and Robert E. Kraut and Laura Dabbish},
  booktitle = {Proceedings of the 2017 ACM Conference on Computer Supported Cooperative Work and Social Computing (CSCW)},
  year = {2017},
  pages = {111--125},
  publisher = {ACM},
  doi = {10.1145/2998181.2998277}
}

@inproceedings{Bryant2005BecomingWikipedian,
  title = {Becoming Wikipedian: Transformation of Participation in a Collaborative Online Encyclopedia},
  author = {Susan L. Bryant and Andrea Forte and Amy Bruckman},
  booktitle = {Proceedings of the 2005 International ACM SIGGROUP Conference on Supporting Group Work (GROUP)},
  year = {2005},
  pages = {1--10},
  publisher = {ACM},
  doi = {10.1145/1099203.1099205}
}

@inproceedings{Lampinen2011DisclosureManagement,
  title = {We're In It Together: Interpersonal Management of Disclosure in Social Network Services},
  author = {Airi Lampinen and Vilma Lehtinen and Asko Lehmuskallio and Sakari Tamminen},
  booktitle = {Proceedings of the SIGCHI Conference on Human Factors in Computing Systems (CHI)},
  year = {2011},
  pages = {3217--3226},
  publisher = {ACM},
  doi = {10.1145/1978942.1979420}
}

@inproceedings{Bernstein2011Fourchan,
  title = {4chan and /b/: An Analysis of Anonymity and Ephemerality in a Large Online Community},
  author = {Michael S. Bernstein and Andr{\'e}s Monroy-Hern{\'a}ndez and Drew Harry and Paul Andr{\'e} and Katrina Panovich and Greg Vargas},
  booktitle = {Proceedings of the 5th International AAAI Conference on Weblogs and Social Media (ICWSM)},
  year = {2011},
  pages = {50--57}
}

@inproceedings{Cheng2017Troll,
  title = {Anyone Can Become a Troll: Causes of Trolling Behavior in Online Discussions},
  author = {Justin Cheng and Michael Bernstein and Cristian Danescu-Niculescu-Mizil and Jure Leskovec},
  booktitle = {Proceedings of the 2017 ACM Conference on Computer Supported Cooperative Work and Social Computing (CSCW)},
  year = {2017},
  pages = {1217--1230},
  publisher = {ACM},
  doi = {10.1145/2998181.2998213}
}

@article{Rashidi2020SanctioningNorms,
  title = {"It's Easier Than Causing Confrontation": Sanctioning Strategies to Maintain Social Norms and Privacy on Social Media},
  author = {Yasmeen Rashidi and Apu Kapadia and Christena Nippert-Eng and Norman M. Su},
  journal = {Proceedings of the ACM on Human-Computer Interaction (PACMHCI)},
  volume = {4},
  number = {CSCW1},
  pages = {23:1--23:25},
  year = {2020},
  doi = {10.1145/3392827}
}

@article{Fiesler2019FanNorms,
  title = {Creativity, Copyright, and Close-Knit Communities: A Case Study of Social Norm Formation and Enforcement},
  author = {Casey Fiesler and Amy S. Bruckman},
  journal = {Proceedings of the ACM on Human-Computer Interaction (PACMHCI)},
  volume = {3},
  number = {GROUP},
  pages = {241:1--241:24},
  year = {2019},
  doi = {10.1145/3361122}
}

@inproceedings{Butler2008WikipediaPolicies,
  title = {Don't Look Now, But We've Created a Bureaucracy: The Nature and Roles of Policies and Rules in Wikipedia},
  author = {Brian Butler and Elisabeth Joyce and Jacqueline Pike},
  booktitle = {Proceedings of the SIGCHI Conference on Human Factors in Computing Systems (CHI)},
  year = {2008},
  pages = {1101--1110},
  publisher = {ACM},
  doi = {10.1145/1357054.1357227}
}

@book{Henderson1991DesignAtWork,
  author    = {Austin Henderson and Morten Kyng},
  title     = {Design at Work: Cooperative Design of Computer Systems},
  year      = {1991},
  publisher = {Lawrence Erlbaum Associates},
  address   = {Hillsdale, NJ}
}

@book{Ericsson1993,
  author    = {Ericsson, K. Anders and Simon, Herbert A.},
  title     = {Protocol Analysis: Verbal Reports as Data},
  year      = {1993},
  publisher = {The MIT Press},
  address   = {Cambridge, MA},
  edition   = {Revised},
  isbn      = {0-262-27239-3}
}

@incollection{maclean2020questions,
  title={Questions, options, and criteria: Elements of design space analysis},
  author={MacLean, Allan and Young, Richard M and Bellotti, Victoria ME and Moran, Thomas P},
  booktitle={Design rationale},
  pages={53--105},
  year={2020},
  publisher={CRC Press}
}

@article{knuth1984literate,
  title={Literate programming},
  author={Knuth, Donald Ervin},
  journal={The computer journal},
  volume={27},
  number={2},
  pages={97--111},
  year={1984},
  publisher={Oxford University Press}
}

@inproceedings{kery2018story,
  title={The story in the notebook: Exploratory data science using a literate programming tool},
  author={Kery, Mary Beth and Radensky, Marissa and Arya, Mahima and John, Bonnie E and Myers, Brad A},
  booktitle={Proceedings of the 2018 CHI conference on human factors in computing systems},
  pages={1--11},
  year={2018}
}

@article{ragan2015characterizing,
  title={Characterizing provenance in visualization and data analysis: an organizational framework of provenance types and purposes},
  author={Ragan, Eric D and Endert, Alex and Sanyal, Jibonananda and Chen, Jian},
  journal={IEEE transactions on visualization and computer graphics},
  volume={22},
  number={1},
  pages={31--40},
  year={2015},
  publisher={IEEE}
}

@inproceedings{goyal2016effects,
  title={Effects of sensemaking translucence on distributed collaborative analysis},
  author={Goyal, Nitesh and Fussell, Susan R},
  booktitle={Proceedings of the 19th ACM Conference on Computer-Supported Cooperative Work \& Social Computing},
  pages={288--302},
  year={2016}
}

@article{catledge1995characterizing,
  title={Characterizing browsing strategies in the World-Wide Web},
  author={Catledge, Lara D and Pitkow, James E},
  journal={Computer Networks and ISDN systems},
  volume={27},
  number={6},
  pages={1065--1073},
  year={1995},
  publisher={Elsevier}
}

@article{pirolli1999information,
  title={Information foraging.},
  author={Pirolli, Peter and Card, Stuart},
  journal={Psychological review},
  volume={106},
  number={4},
  pages={643},
  year={1999},
  publisher={American Psychological Association}
}

@inproceedings{dabbish2012social,
  title={Social coding in GitHub: transparency and collaboration in an open software repository},
  author={Dabbish, Laura and Stuart, Colleen and Tsay, Jason and Herbsleb, Jim},
  booktitle={Proceedings of the ACM 2012 conference on computer supported cooperative work},
  pages={1277--1286},
  year={2012}
}

@inproceedings{mamykina2011design,
  title={Design lessons from the fastest q\&a site in the west},
  author={Mamykina, Lena and Manoim, Bella and Mittal, Manas and Hripcsak, George and Hartmann, Bj{\"o}rn},
  booktitle={Proceedings of the SIGCHI conference on Human factors in computing systems},
  pages={2857--2866},
  year={2011}
}
